\documentclass[reprint, superscriptaddress]{revtex4-2}

\usepackage[toc,page]{appendix}
\usepackage[british]{babel}
\usepackage[utf8]{inputenc}
\usepackage{amsmath,amssymb}
\usepackage{graphicx}
\usepackage{ae}
\usepackage{esdiff}
\usepackage{braket}
\usepackage{bbm}
\usepackage{simplewick}
\usepackage{float}
\usepackage{braket}
\usepackage{simplewick}
\usepackage{color}
\usepackage{framed}
\usepackage[caption=false]{subfig}
\usepackage[pdftex,colorlinks=true,linkcolor=blue,citecolor=blue,urlcolor=black]{hyperref}
\usepackage{mathtools}
\usepackage{enumitem}
\usepackage{ragged2e}
\usepackage{floatflt}
\usepackage{import}
\usepackage{titlesec}
\usepackage{multirow}
\usepackage{etoolbox}
\usepackage{appendix}
\usepackage{dsfont}
\usepackage{comment}
\usepackage{xcolor}
\usepackage{blindtext}
\usepackage[makeroom]{cancel}

\begin{document}

\newcommand{\norm}[1]{\left\lVert#1\right\rVert}

\def \r{{\boldsymbol{r}}}
\def \k{{\boldsymbol{k}}}
\def \p{{\boldsymbol{p}}}
\def \q{{\boldsymbol{q}}}
\def \x{{\textbf{x}}}
\def \A{{\textbf{{A}}}}
\def \a{{\textbf{{a}}}}
\def \b{{\textbf{{b}}}}
\def \c{{\textbf{{c}}}}
\def \z{{\textbf{{z}}}}
\def \0{{\boldsymbol{{0}}}}
\def \dl{\frac{\partial}{\partial l}}
\def \P{{\boldsymbol{P}}}
\def \K{{\boldsymbol{K}}}
\def \sigmad{{\sigma_{\downarrow\uparrow}}}
\def \uone{{\boldsymbol{u}_1}}
\def \utwo{{\boldsymbol{u}_2}}
\def \edown{{\epsilon_{\downarrow}}}
\def \omfl{{\omega_{\text{FL}}}}
\def \piph{\Pi_\text{ph}}
\def \sign{ \text{sign}}
\def \lamt{\tilde{\lambda}}
\def \Sigmab{\Sigma_{\omega^2, \text{bs}}}
\def \intk{{\int_\textbf{k}}}
\def \eone{{\epsilon_{\uparrow,1}}}
\def \etwo{{\epsilon_{\uparrow,2}}}
\def \Eone{{\epsilon_{\downarrow,1}}}
\def \Ims{\text{Im} \left[ \Sigma(\omega) \right]}

\definecolor{mgrey}{RGB}{63,63,63}
\definecolor{mred}{RGB}{235,97,51}
\newcommand{\mg}[1]{{\color{mgrey}{#1}}}
\newcommand{\mr}[1]{{\color{mred}{#1}}}

\newcommand{\beq} {\begin{equation}}
\newcommand{\eeq} {\end{equation}}
\newcommand{\bea} {\begin{eqnarray}}
\newcommand{\eea} {\end{eqnarray}}
\newcommand{\be} {\begin{equation}}
\newcommand{\ee} {\end{equation}}
\newcommand{\red}[1]{{\color{red}{#1}}}

\def\BigColSep{\setlength{\arraycolsep}{50pt}}

\title{One-dimensional scattering of two-dimensional fermions near quantum criticality}

\author{Dimitri Pimenov}\email{dpimenov@umn.edu}
\author{Alex Kamenev}
\author{Andrey V.\ Chubukov}
\affiliation{William I. Fine Theoretical Physics Institute, University of Minnesota, Minneapolis, MN 55455, USA}
\begin{abstract}
Forward and backscattering  play an exceptional role in the physics of two-dimensional
interacting fermions.
 In a Fermi liquid,  both give rise to a  non-analytic
$\omega^2 \ln(\omega)$  form of the fermionic scattering rate at second order in the interaction.
Here we argue that higher powers of $\ln(\omega)$  appear
in the backscattering contribution
 at higher orders.  We show that these terms come from  ``planar'' processes, which are effectively one-dimensional.  This is explicitly  demonstrated  by
   extending  a Fermi liquid to the limit
   of $N \gg 1$ fermionic flavors, when only planar processes survive.
   We sum the leading logarithms for the case  of a 2D Fermi liquid near a nematic transition and obtain an expression for the scattering rate at $T=0$  to all orders in the interaction.   For a repulsive interaction, the resulting rate is logarithmically suppressed, and the   result
     is valid down to $\omega = 0$. For  an  attractive interaction, the ground state is an $s$-wave superconductor with a gap $\Delta_0$. We show that in this case the scattering rate increases
      as $\omega$ is reduced towards $\Delta_0$.
       At $\omega \geq \Delta_0$, the behavior of the scattering rate  is rather unconventional as
   many pairing channels compete near a nematic critical point,  and $s$-wave  wins only by a narrow margin. We take superconductivity into consideration and obtain the scattering rate also at smaller $\omega     \simeq \Delta_0$.
\end{abstract}
 \maketitle
\section{Introduction}

Quantum critical metals (QCM) constitute a rich and ever-evolving field of modern condensed matter physics. The broad interest in their behavior is fueled not only by the increasing number of experimentally probed systems,
 such as cuprates and iron-based materials \cite{PhysRevB.81.184519, hussey2008phenomenology, doi:10.1146/annurev-conmatphys-031119-050558,RevModPhys.73.797}, but also by the advent of modern numerical techniques, in particular Quantum Monte-Carlo \cite{berg2012sign, PhysRevX.6.031028, Lederer4905, doi:10.1146/annurev-conmatphys-031218-013339}. One exciting aspect of QCMs is their prototypical non-Fermi liquid behavior, and it is a theorist's dream to determine the associated critical  behavior exactly.
 So far, this has only been achieved in the special cases  
 such as the SYK model \cite{PhysRevLett.70.3339, Kitaev2015}, a particular matrix large $N$ theory~\cite{PhysRevLett.123.096402}, a scalar large $N$ model with a particular dispersion of a critical boson \cite{PhysRevB.82.045121}, and models obtained by dimensional regularization \cite{PhysRevB.88.245106}.
  Besides, a fully  self-consistent description of a spin density wave quantum critical point (QCP) in (2+1)D at the lowest energies has also been  proposed in Ref.\ \cite{PhysRevX.7.021010} and Ref.~\cite{PhysRevB.90.045121} presented
 the general description of critical theories with a chiral Fermi surface. 

A {particular example of a} QCM, which defies complete understanding despite years of study \cite{POLCHINSKI1994617, PhysRevB.50.14048, PhysRevB.50.17917, PhysRevB.64.195109, PhysRevLett.91.066402, PhysRevB.73.045127, PhysRevB.73.085101, PhysRevB.74.195126,  doi:10.1146/annurev-conmatphys-070909-103925,  metlitski2010quantum, holder2015anomalous, PhysRevB.82.045121, PhysRevB.88.245106,PhysRevB.96.155125, PhysRevB.103.235129},
is the one at an Ising nematic QCP in (2+1)D.
Here, the fermionic self-energy both at one and two loop level scales as $\omega^{2/3}$, which is a hallmark of a non-Fermi liquid. However, the loop expansion lacks a control parameter. It was originally   believed \cite{POLCHINSKI1994617,PhysRevB.50.14048} that one can justify the expansion by extending the model to  $N$ fermion flavors   and taking the limit $N \to \infty$. Indeed, at large $N$, the prefactor for $\omega^{2/3}$ at the two-loop order is parametrically smaller than the one-loop result  \cite{PhysRevB.50.14048,PhysRevB.74.195126}. However, it was later recognized by S.-S.\ Lee \cite{PhysRevB.80.165102} that this does not hold  beyond the two-loop order. He analyzed a simplified ``one-patch'' model  without backscattering  and  explicitly demonstrated that planar diagrams, which emerge at three-loop and higher orders, are not small in $1/N$.  Subsequent studies of  the Ising nematic  model within the ``two-patch theory'', which includes backscattering, found more contributions from planar diagrams, not small in $1/N$, and discovered  additional logarithmic singularities (Refs.~\cite{metlitski2010quantum, holder2015anomalous}).

There are several equivalent characterizations of planar diagrams, which can all be invoked to extract their
special behaviour at large $N$. The original name-giving criterion in the high-energy literature \cite{HOOFT1974461} is that they can be drawn on a plane without holes.  From the condensed matter perspective, the planar diagrams
describe a subset of scattering processes which have a (1+1)-dimensional structure, despite the original problem being (2+1)D -- more specifically, in these diagrams  the curvature term in the fermionic dispersion cancels out \cite{metlitski2010quantum},\footnote{We note that the two-dimensionality still shows up via Landau damping, without it the fermionic self-energy would be more singular than $\omega^{2/3}$ \cite{PhysRevB.73.085101, PhysRevLett.97.226403} }.

Because the leading one-loop self-energy $\omega^{2/3}$  also comes from (1+1)D processes
\cite{PhysRevLett.95.026402, PhysRevB.73.045128}, the $N \to \infty$ limit at a QCP should be described by some effective (1+1)-dimensional theory.  However, the detailed structure of higher-loop planar diagrams is not yet known, and the story is further complicated by apparent UV-divergences at higher-loop orders~\cite{metlitski2010quantum,holder2015anomalous}.  As a result, the form of fermionic and bosonic propagators at a QCP in the large $N$ limit remains unclear.

In this paper we analyze the scattering rate in a (2+1)D Fermi liquid away from a nematic
  QCP.   Our first goal is to understand the  role of planar diagrams in a Fermi liquid regime. In a generic  2D
   Fermi liquid, the imaginary part of the one-loop self-energy $\text{Im}\left[\Sigma (\omega)\right]$ at $k=k_F$ scales as $\omega^2 \ln(\omega)$  (as $ (\omega^2 + (\pi T)^2)\ln\left[{\text{max}} \left(\omega, T\right)\right]$ at a finite temperature $T$). This non-analytic form comes from forward and backscattering processes~\cite{PhysRevB.71.205112},
 which can be regarded as effectively one-dimensional. We analyze $\text{Im}\left[\Sigma (\omega)\right]$ at $T=0$ beyond one-loop order, and show that it contains a series of higher powers of $\ln(\omega)$. The logarithms come from the Cooper part of backscattering, which is manifestly (1+1)D,  and we argue that these terms can be equally viewed as coming from planar processes.  We demonstrate this explicitly by analyzing the system behavior in the  limit of large $N$, when only planar processes survive. We explicitly  sum up series of logarithms and obtain the scattering rate to logarithmic accuracy to all orders in the interaction. We show, however, that in a Fermi liquid not all planar processes are (1+1)D -- the ones that do not contain the highest power of $\ln (\omega)$ at any loop order involve scattering with deviations from a single direction by momenta $q \sim M$, where $M^2$  scales with  the distance from a QCP \footnote{At a QCP, all planar processes become (1+1)D, and, simultaneously, the argument of the logarithm becomes of order one. More specifically, it becomes
a function of the combination $q^3/\omega$, where $q$ and $\omega$ are relevant internal momenta and frequency in a
planar diagram. Both are infinitesimally small at vanishing external frequency, and typical $q^3$ are of order $\omega$.Then the effective dimension of the full $N = \infty$ theory becomes (1+1)D, with no separation into the leading and subleading terms.}.

For a toy model with repulsive interaction in the particle-particle channel, the full $\Ims$ to logarithmic accuracy
remains finite and is reduced, roughly by a half, compared to the one-loop expression. A similar result holds for the $T^2$
term in the specific heat of a 2D Fermi gas~\cite{PhysRevB.74.075102, PhysRevB.76.165111}.
 For the realistic case of an attractive pairing interaction,
 the ground state is  an $s$-wave superconductor~\cite{PhysRevLett.77.3009,Moon2010,PhysRevLett.114.097001, PhysRevB.90.165144, PhysRevB.90.174510, PhysRevB.91.115111,PhysRevB.94.115138, CHOWDHURY2020168125}.
 We show that, in this case, the prefactor of the $\omega^2$ term in  $\Ims$ increases with $\ln(\omega)$, and the summation of the logarithms breaks down at some energy scale $\omega  \geq \Delta_0$, where $\Delta_0$ is a pairing gap.

Our second goal is then to correctly incorporate the superconducting ground state and obtain $\Ims$
for $\omega \simeq \Delta_0$. This is of relevance both for experiments and for
Quantum Monte-Carlo studies, which found~\cite{PhysRevX.6.031028,Lederer4905} that superconductivity must be included to understand the data for the self-energy at the lowest frequencies/temperatures.
 We argue that in a 2D Fermi liquid, the frequency range where logarithmic renormalizations from planar diagrams are relevant, but superconductivity can be neglected, can be made wide even near a QCP,
  by restricting to a  weak coupling. We show that in this case one can incorporate superconductivity in a
     controllable way at $\omega \simeq \Delta_0$ \footnote{Exactly at a QCP, the scale $\Delta_0$ is of the same order as the upper edge of the non-Fermi-liquid~\cite{CHUBUKOV2020168142}, and there is no sizable frequency range for the renormalization of the self-energy by planar diagrams without including superconducting fluctuations, unless one makes specific assumptions about the shape of the Fermi surface \cite{PhysRevB.80.165102}.}.

The fermionic self-energy at $\omega \simeq \Delta_0$ has been analyzed in the past for a  conventional $s$-wave superconductor  with momentum-independent  pairing interaction  ~\cite{PhysRevB.42.10211,Larkin2008}.    We show that our case is qualitatively different, because the interaction, mediated by soft nematic fluctuations, gives rise to near-equal attraction in a large number of pairing channels, and $s$-wave is only slightly preferable
 over other pairing symmetries ~\cite{PhysRevLett.114.097001, PhysRevB.98.220501}. In this case, we show that $\Ims$ has the same form as in a pure $s$-wave superconductor between $\Delta_0$ (below which $\Ims = 0$) and slightly over $3\Delta_0$, where the system realizes that attraction in the $s$-wave channel is  larger than that in  other channels.
At larger $\omega$,   all pairing channels contribute equally, and  the form of $\Ims$ changes qualitatively.

 The
 paper is structured as follows: In Sec.\ \ref{modelsec}, we present the
model, recapitulate one-loop results for $\Sigma[\omega]$  and discuss the relevant energy scales. In Sec.\ \ref{planarnonplanarsec}, we analyze the structure of three-loop planar diagrams and demonstrate that the theory is effectively one-dimensional at logarithmic order, but not beyond. In the next Sec.\ \ref{flselfsecrep}, we sum up the series of the leading $\omega^2 (\ln{\omega})^n$ contributions to the self-energy from backscattering at all loop orders.
In Sec.\ \ref{repulsive}, we consider a toy model with repulsive pairing interaction and show that the summation of the leading logarithms converges at any frequency.
The realistic case of an attractive interaction is analyzed in Sec. \ref{attractivenormalstatesec}, where we
show that the convergence only holds  up to  $\omega  \sim \Delta_0$. In Sec.\ \ref{scscatesec}, we demonstrate how the results get modified once we add superconductivity and obtain the expressions for $\Ims$ at $\omega \simeq \Delta_0$. Conclusions and an outlook are presented in Sec.\ \ref{concsec}. Various technical details are relegated to the Appendices.

\section{Model and one-loop results}
\label{modelsec}

\begin{figure}
\centering
\includegraphics[width=\columnwidth]{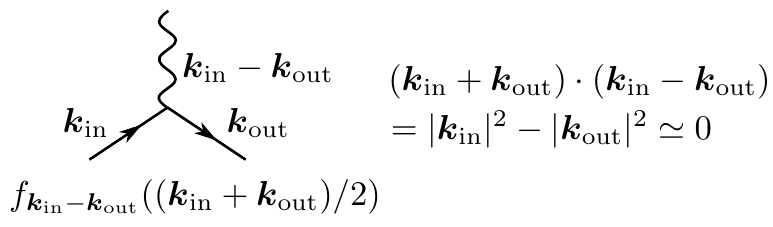}
\caption{{Definition of form factor, see main text}}
\label{formfactor}
\end{figure}

We consider a system of interacting $N$-flavor fermions in two dimensions  at $T=0$.
Two spin projections are incorporated  into $N$ such that the physical case of spin-1/2  fermions corresponds to $N = 2$. We assume that the system is {isotropic and} close to a QCP towards a nematic order {($d$-wave Pomeranchuk instability)} and that near a QCP the dominant interaction between fermions is mediated by soft fluctuations of a nematic order parameter.  The corresponding $T = 0$ Euclidean action \cite{PhysRevLett.91.066402, PhysRevB.73.045127} is given by
\begin{align}
\label{mainaction}
\mathcal{S} &= \mathcal{S}_0 + \mathcal{S}_{\text{int}} \ , \\ \notag
\mathcal{S}_0 &= \sum_{\sigma = 1}^N \int_{k} \bar\psi_\sigma(k) (-i\omega_m + \xi_\k)\psi_\sigma(k)\  ,  \\ \notag
\mathcal{S}_\text{int} &=
\frac{g}{V}  \sum_{\sigma, \sigma^\prime} \int_{k,p,q}  D_0(\q)
f_\q(\k) f_\q(\p)  \times  \\ & \notag \quad \bar\psi_\sigma(k + q/2) \bar\psi_{\sigma^\prime} (p - q/2) \psi_{\sigma^\prime}(p+q/2) \psi_\sigma(k-q/2) \ , \\  \notag
\int_k &\equiv \int \frac{d\omega_m}{2\pi} \frac{d\k}{(2\pi)^2} \ .
\end{align}
Here $\xi_\k = v_F(|\k| - k_F)$ is the flavor-independent fermionic dispersion, linearized around the Fermi momentum,
$\omega_m$ is a Matsubara frequency, $g$ is  the coupling,  which   we set to be small compared to the Fermi energy, $g/E_F \ll 1$, and
\begin{align}
D_0(\q) = - \frac{1}{q^2 + M^2} \ , \quad q \equiv |\q| \ ,
\end{align}
is the bare bosonic propagator. The parameter $M$ (bosonic mass) measures the deviation from the quantum critical point, and has units of momentum. Near a QCP,
\begin{align}
\epsilon \equiv \frac{M}{k_F}  \ll 1 \ .
\end{align}
{Finally, $f_\q(\k) = \cos(2 \theta_{\k\q})$ (with $ \theta_{\k\q} = \measuredangle(\k,\q))$ is the $d$-wave nematic form factor (not related to pairing symmetry), which specifies the symmetry of the ordered  state. At a general boson-fermion vertex with incoming and outgoing fermionic momenta $\k_{\text{in}}, \k_\text{out}$, the half-sum $(\k_{\text{in}} + \k_{\text{out}})/2$ corresponds to $\k$ in the above defintion of $f$, while $\k_\text{out} - \k_\text{in}$ corresponds to $\q$. In our analysis, all momenta will fulfill $|\k_{\text{in}}|, |\k_\text{out}| \simeq k_F$, resulting in $\k \perp \q$ and  $f \simeq - 1$. (see Fig.\ \ref{formfactor}). Since all relevant expressions contain $f^2$, we can set $f = 1$ from the start. The form factor is important at high energies of order $\mathcal{O}(E_F)$ or in the ordered state only, which are not considered in this paper. For a short discussion of lattice effects, see Sec.\ \ref{concsec}. }

\begin{figure}
\centering
\includegraphics[width=.5\columnwidth]{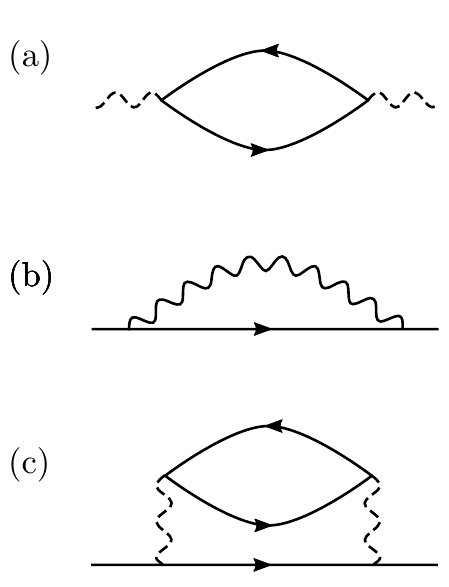}
\caption{One-loop diagrams. (a) Particle-hole bubble $\Pi_{\text{ph}}$, which renormalizes the bare interaction $D_0$ (dashed wavy lines). (b) One-loop self-energy $\Sigma^{(1)}$. Full wavy line represents the renormalized interaction $D$. (c) Diagram which effectively determines $\Sigma^{(1)}(\omega)$ for $\omega \ll \omfl$ (interaction lines are static).}
\label{diagtable}
\end{figure}

The action of Eq.\ (\ref{mainaction}) is obtained from a microscopic Hamiltonian with 4-fermion interaction by integrating out fermions with energies above a certain cutoff. To account for screening from fermions with energies smaller than the cutoff, the bare propagator  $D_0(\q)$ should be dressed by a particle-hole bubble $\Pi_{\text{ph}}$, see Fig.\ \ref{diagtable}(a) and  App.\ \ref{bubblesapp}. The dressed propagator $D(\omega_m,\q)$ is
\begin{align}
\label{Landaudampingdef}
&D(\omega_m,\q) = - \frac{1}{q^2 + M^2 + g \Pi_{\text{ph}} (\omega_m, \q) }
 , &\\ \notag
 &{\text {where}}  \quad
  \Pi_{\text{ph}} (\omega_m , \q) = N\rho \frac{|\omega_m|}{\sqrt{\omega^2_m + (v_F q)^2}} \ .  &
\end{align}
$\rho = m/(2\pi)$ the density of states per flavor, and $m = k_F/v_F$ the fermionic mass.
   We assume, following earlier works \cite{metlitski2010quantum, doi:10.1080/0001873021000057123}, that the
static part of $\Pi_{\text{ph}}$ is already incorporated into $D(\omega_m, \q)$.
   At $\omega_m \ll v_F q$, which will be relevant to our analysis, $\Pi_{\text{ph}} (\omega_m , \q) \approx N \rho |\omega_m|/(v_F q)$ describes (in real frequencies) the Landau damping of
 a boson due to a decay into the particle-hole continuum.

\begin{figure}[t]
\centering
\includegraphics[width=\columnwidth]{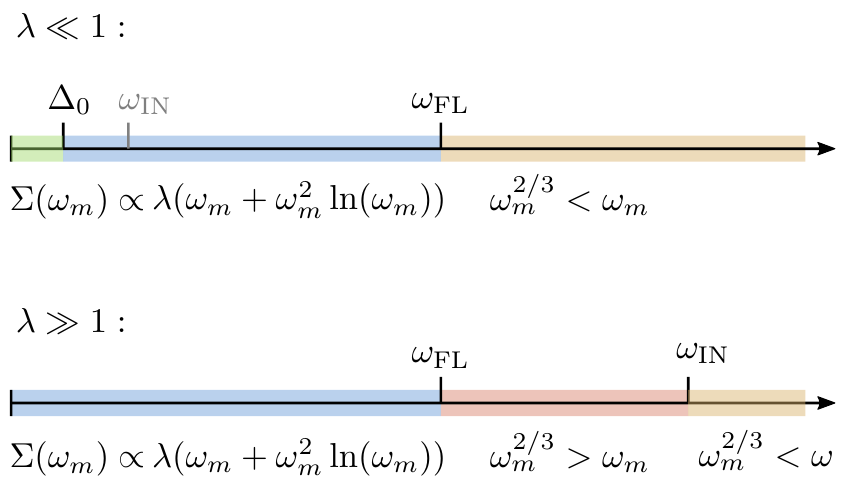}
\caption{Energy scales. For $\lambda \ll 1$, the self-energy is always smaller than the bare frequency, $\Sigma(\omega_m)< \omega_m$, and there is no true non-Fermi-liquid. The green sliver marks the superconducting region for attractive interactions discussed in Sec.\ \ref{scscatesec}, with $\Delta_0 \simeq \omfl \exp(-\lambda)$. For $\lambda \gg 1$, a true non-Fermi-liquid region (red) develops where  $\Sigma(\omega_m)> \omega_m$. Superconductivity in the strong coupling case is outside the scope of this paper. The scale $v_F M > \omfl$ is not explicitly shown. }
\label{scales}
\end{figure}

With these ingredients, we can evaluate the fermionic self-energy for the external momentum $\k$ on the Fermi surface,
$\Sigma(\omega_m,\k_F) \equiv  \Sigma(\omega_m)$
$(\text{with}\  G^{-1} = G^{-1}_0 - \Sigma$). The one-loop  self-energy  $\Sigma^{(1)}(\omega_m)$, represented by the diagram in Fig.\ \ref{diagtable}(b), has been obtained before, see, e.g., Refs.\ \cite{PhysRevB.74.195126, metlitski2010quantum},
and we just present the result:
\begin{align}
\label{sigma1resultmain}
& \Sigma^{(1)}(\omega_m) \simeq \\& \notag  \begin{dcases} - i
\lambda  \omega_m +  i \text{sign}(\omega_m) \frac{2\lambda}{\pi}
\frac{\omega^2_m}{\omfl} \ln\left(\frac{\omfl}{|\omega_m|}\right), \! &\omega_m \ll \omfl \\
\omega_{\text{IN}}^{1/3}  \omega^{2/3}_m, \! &\omega_m \gg \omfl.
\end{dcases}
\end{align}
The first line in Eq.\ \eqref{sigma1resultmain} has a familiar Fermi-liquid form
 in 2D. It contains
two  parameters
\begin{align}
\label{lambdadef}
\lambda \equiv \frac{g}{4 \pi v_F M} , \quad \quad \omfl  \equiv \frac{ v_F M^3}{N g \rho}  \ .
\end{align}
Here, $\lambda$ is an effective coupling, which will be used to control perturbation theory, and $\omfl$
  serves as UV cutoff for the logarithms and also marks the crossover from a Fermi-liquid to a non-Fermi-liquid regime.
  The second line  in \eqref{sigma1resultmain}  is the  one-loop formula for the self-energy in the non-Fermi liquid regime.
 The scale
 \beq
\label{win}
\omega_{\text{IN}} = \frac{8}{3\sqrt{3}} \lambda^3 \omfl \
 \eeq
 is the upper boundary of the true non-Fermi-liquid region, where $\Sigma(\omega_m) > \omega_m$.
  The relevant energy scales are summarized in Fig.\ \ref{scales} for both $\lambda \ll 1$, $\lambda \gg 1$.

In deriving  Eq.\ \eqref{sigma1resultmain} we assumed that typical internal frequencies $\omega^\prime_m$ and momenta $q$ satisfy
$\omega^\prime_m \ll v_F q$. This condition is naturally satisfied at a QCP, where typical $\omega^\prime_m \sim q^{3}$, see e.g.\ \cite{PhysRevX.10.031053}, but in a weakly coupled Fermi liquid at $\lambda \ll 1$ we must additionally require  $\lambda > \epsilon/N$, such that $\omfl
 \leq v_F M$.

For $\omega \ll \omfl$, the Landau damping is small compared to the mass term $M^2$ in $D(\omega_m,\q)$, and the Fermi-liquid form of $\Sigma^{(1)}(\omega_m)$  is obtained by expanding in the Landau damping. In effect, this reduces the evaluation of the diagram in Fig.\ \ref{diagtable}(b) to the evaluation of the two-loop diagram with static interactions $D_0(\q)$ of Fig.\ \ref{diagtable}(c).  Let us now zoom into  the $\omega^2$-part of the Fermi-liquid self-energy at these frequencies:
 \begin{align}
\label{sigmaomegaqdef}
\Sigma^{(1)}_{\omega^2} (\omega_m) \equiv  i \text{sign}(\omega_m)
\frac{2\lambda}{\pi} \frac{\omega^2_m}{\omfl} \ln\left(\frac{\omfl}{|\omega_m|}\right).
\end{align}
 Upon analytical continuation $i\omega_m  \rightarrow \omega + i\delta$,  it maps to
the imaginary part of the retarded self-energy
\begin{align}
\text{Im}\left[\Sigma^R(\omega)\right] =  -\frac{2 \lambda}{\pi} \frac{\omega^2}{\omfl} \ln\left(\frac{\omfl}{|\omega|}\right),
\label{a_1}
\end{align}
which determines the fermionic scattering rate.

\begin{figure}[b]
\centering
\includegraphics[width=\columnwidth]{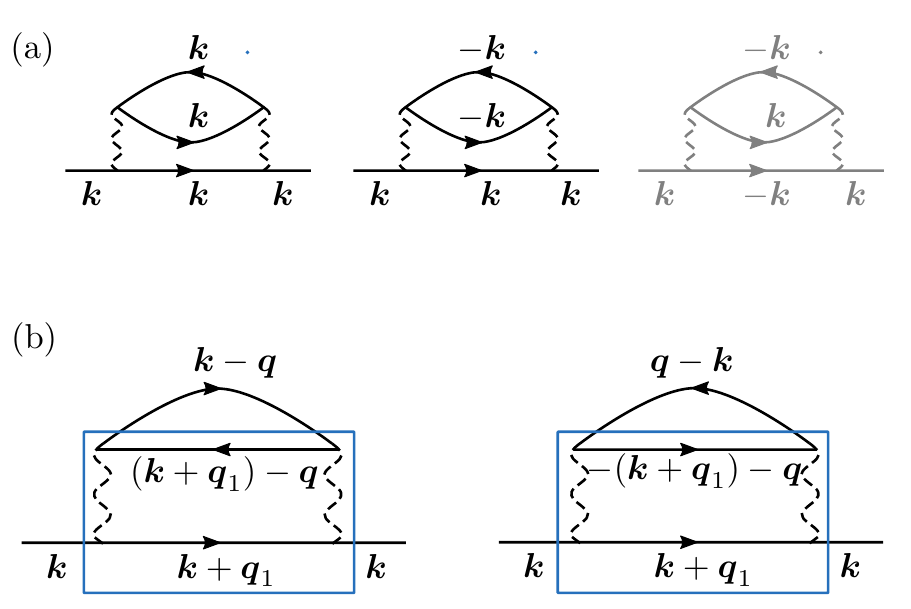}
\caption{One-dimensional scattering at two-loop level. (a) Most singular momentum points (forward and backscattering) which contribute to the logarithm in $\Sigma_{\omega^2}^{(1)}$. The grayed out diagram is also part of backscattering, but its contribution can be neglected since it involves a large momentum transfer $\sim 2k_F$. (b) Way to extract the logarithm from the forward and backscattering processes (see main text).}
\label{oneloop1ddiag}
\end{figure}

For our future considerations it is important that the $\omega_m^2 \ln{(\omega_m)}$ dependence in  Eq.~(\ref{sigmaomegaqdef}) can be traced to effectively one-dimensional scattering, embedded in two dimensions (see Appendix \ref{oneloopfermiapp} for technical details).  Namely,  the processes that contribute to Eq.\ (\ref{sigmaomegaqdef}) are forward scattering and backscattering, with momentum deviations  from $\k$ and ${-\k}$  on the order of external $\omega_m$  (see Fig.~\ref{oneloop1ddiag}(a)). As sketched in Fig.\ \ref{oneloop1ddiag}(b), instead of evaluating the closed particle-hole bubble in the diagram one can explicitly obtain one half of $\omega_m^2 \ln{(\omega_m)}$
 from either of these two  scattering processes (the first diagram describes forward scattering and the second one backscattering). In each case,  the integration over the internal momentum $\q_1$ (assuming $|\q_1| \ll k_F$) and corresponding frequency in the blue boxes in Fig.\ \ref{oneloop1ddiag}(b) produces a Landau damping term $\sim \omega_m^\prime/q$ as the leading dynamical contribution,  where $q = |\q|$ and $\omega_m^\prime$  are the momentum and the frequency of the remaining internal fermion in the diagrams in Fig.\ \ref{oneloop1ddiag}. The integral over the angle between $\k,\q$ (or, equivalently, the component of $\q$ parallel to $\k$) restricts internal frequencies to $\omega_m^\prime \in (0,\omega_m)$, and each diagram yields
\begin{align}
\label{schematic1}
\Sigma_{\omega^2}^{(1)} (\omega_m) \propto \int_0^{\omega_m} d\omega_m^\prime \int_{\omega^\prime_m/v_F}^{\omfl/v_F} dq \frac{\omega_m^\prime}{q} \propto \omega_m^2 \ln(\omega_m) \ ,
\end{align}
as in Eq.~\eqref{sigmaomegaqdef}.

Relevant $q$ in the integral are of order $\omega'_m/v_F$ (to logarithmic accuracy), and relevant $\omega'_m$ are of order of external $\omega_m$, which we set to be the smallest scale in the problem. Then relevant $q^2$ are much smaller than $M^2$,  and hence  the static part of $D(\omega_m, q)$ can be approximated by $1/M^2$, i.e., at this order  static interaction can be treated as momentum-independent. This explains why the prefactor in Eq.~\eqref{sigmaomegaqdef} is $ \lambda/\omfl \propto 1/M^4$ .

\section{Structure and effective dimensionality of higher order diagrams}
\label{planarnonplanarsec}

There are  multiple  self-energy diagrams at higher
 loop  orders, but we have two tuning knobs to single out the most important ones: small external $\omega_m$ and large $N$. Let us focus on large $N$ first. At the QCP, the diagrams with the highest power of $N$ are the planar ones, as discussed in Refs.~\cite{PhysRevB.73.045128, metlitski2010quantum,PhysRevB.82.045121}. We  show   below  that the same holds away from a QCP. We discuss the general structure of planar diagrams in Appendix \ref{planarstructureapp}.

We compute the self-energy order-by-order in the loop expansion on the Matsubara axis, and then convert the
 result onto the real axis.  We associate the full dynamical $D(\omega_m, \q)$ to each bosonic propagator in the diagrams and absorb the $-i\lambda \omega_m$ term in the self-energy into the bare fermionic propagator.

\begin{figure}
\centering
\includegraphics[width=.85\columnwidth]{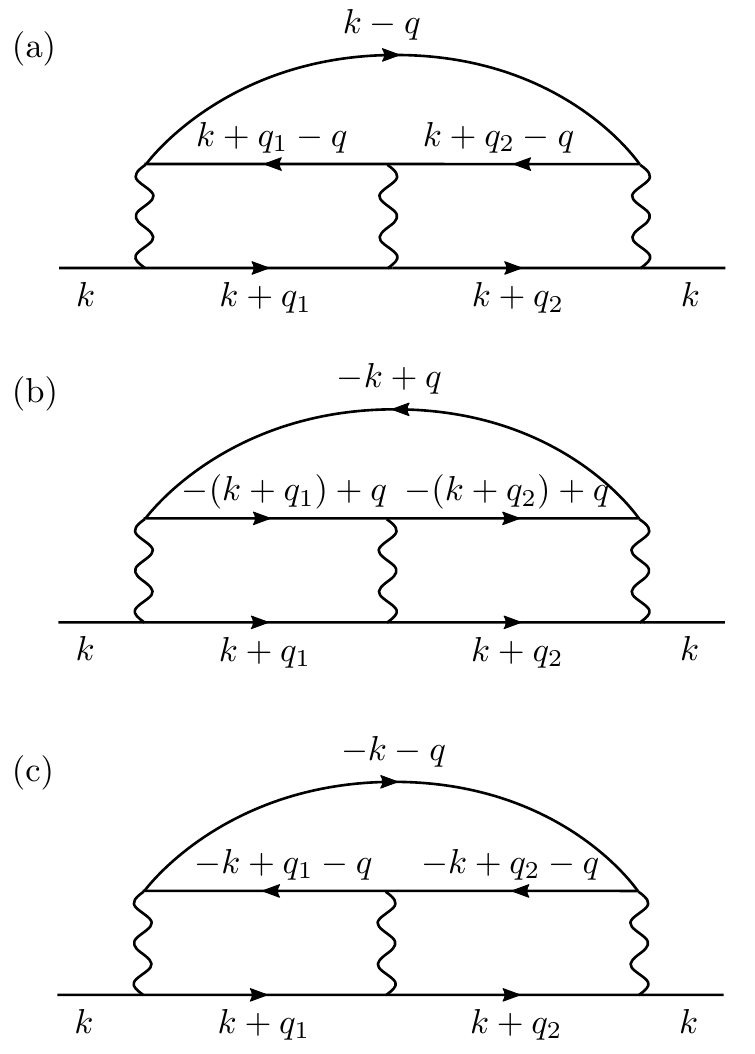}
\caption{Three-loop diagrams. Labels indicate four-momenta (a) Planar diagram in the forward scattering channel (b) Planar diagram in the  backscattering channel (c) Non-planar diagram. }
\label{planar3loop}
\end{figure}

The first planar diagrams beyond one-loop appear at the three-loop order. There are two such diagrams, we show them in Figs. \ref{planar3loop}(a),(b).  Each diagram can be computed explicitly (App \ref{threeloopApp}), and the result is
 (for $\omega_m>0$)
 \begin{align}
\label{forwardresult3loop}
\Sigma^{(3)}_{\omega^2, \text{a}}(\omega_m) &=
i  \frac{1}{\pi^2} \frac{\lambda^2}{1+\lambda} \frac{\omega^2_m}{\omfl}\times(0.56329\hdots)\  ,
\\
\label{backscatteringhere}
\Sigma^{(3)}_{\omega^2,\text{b}}(\omega_m) &= i \frac{1}{2\pi} \frac{\lambda^2}{1+\lambda} \frac{\omega^2_m}{\omfl} \ln^2\left(\frac{\omfl}{\omega_m}\right) \ .
\end{align}
Comparing these two terms with the one-loop result, Eq.~(\ref{sigmaomegaqdef}), we find that the power of $N$ is the same, and there is an additional factor $\lambda/(1+\lambda)$. Besides,  $\Sigma^{(3)}_{\omega^2,\text{a}}( \omega_m)$ contains no logarithm, while $\Sigma^{(3)}_{\omega^2,\text{b}}(\omega_m)$ contains $\ln^2{(\omega_m)}$.

For comparison, consider one of the non-planar diagrams,  like the one in Fig.\ \ref{planar3loop}(c). Evaluating the corresponding integrals, we obtain
\begin{align}
\label{backresult3loop_1}
\Sigma^{(3)}_\text{c}(\omega_m) =
- i \frac{2}{N} \frac{\lambda}{(1+\lambda)\pi} \frac{\omega^2_m}{\omfl} \ln\left(\frac{\omfl}{\omega_m}\right)  \ .
\end{align}
Comparing this result with Eq.~(\ref{sigmaomegaqdef}), we see that $\Sigma^{(3)}_\text{c}(\omega_m)$ contains an additional factor $1/N$.

\begin{figure}
\centering
\includegraphics[width=.85\columnwidth]{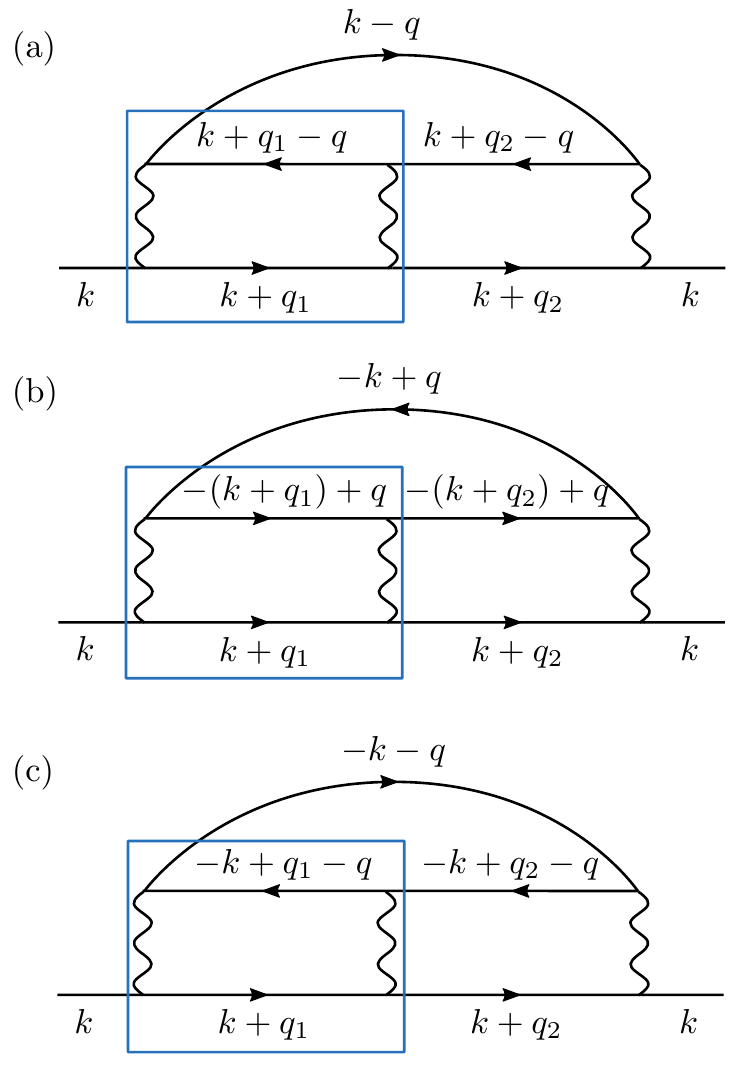}
\caption{Three-loop diagrams as in Fig.\ \ref{planar3loop}, with the dressed vertices marked by blue boxes.}
\label{planar3loopvertices}
\end{figure}

We now look more closely at the two planar diagrams and explore another knob: small external $\omega_m$.
We remind the reader that  the $\omega^2_m \ln{(\omega_m)}$  term in the one-loop self-energy,
 Eq.~(\ref{sigmaomegaqdef}),  comes from forward scattering and backscattering.
 The issue we
want to address is  whether the self-energies $\Sigma^{(3)}_{\omega^2,\text{a}}(\omega_m)$ and $\Sigma^{(3)}_{\omega^2,\text{b}}(\omega_m)$ can be cast into the same form as Eq.~(\ref{sigmaomegaqdef})
with dressed static forward scattering and backscattering amplitudes. Simple experimentation shows that this holds if  two of the three internal momenta and frequencies in Fig.~\ref{planar3loop} (a),(b), (either $q$ and $q_1$, or $q$ and $q_2$,  where $q = ( \omega'_m, \q)$), scale with $\omega_m$ and hence are infinitesimally small at $\omega_m \to 0$. At a QCP,  $\omega_m$ is the only scale in the problem, and all internal momenta/frequencies in the planar diagrams are necessary infinitesimally small at $\omega_m \to 0$. However, in a Fermi liquid there exists an internal momentum scale $M$ (energy scale $v_F M$), and some internal momenta may be of order $M$. We illustrate this  in Fig.~\ref{planar3loopvertices}, where a blue box labels a  dressed vertex.

The computation of the dressed vertices at vanishing external $q$ and $q_2$ is straightforward and we skip the details. For the planar diagram of Fig.~\ref{planar3loopvertices}(b),  we find that,
 to logarithmic accuracy, the result can be  expressed via the dressed  static backscattering amplitude, i.e., the sum of the backscattering part of  $\Sigma^{(1)}_{\omega^2}(\omega_m)$  and    $\Sigma^{(3)}_{\omega^2,\text{b}}(\omega_m)$ can be re-expressed as
  \begin{align}
\label{sigmaomegaqdef_new}
&\Sigma_{\omega^2, \text{bs}}(\omega_m)  \equiv \\ & \notag  i \text{sign}(\omega_m)
\frac{2\lambda}{\pi \omfl}  \int_0^{|\omega_m|} d\omega_m^\prime\omega_m^\prime \int_{\omega_m^\prime}^{\omfl} \frac{dz}{z} \left[\frac{M^2}{g}\Gamma_{\text{bs}} (z)\right]^2 ,
\end{align}
where
\beq
\Gamma_{\text{bs}} (z) = - \frac{g}{M^2} \left[1 + \frac{1}{2} 
 \lambda Z  \ln\left(\frac{\omfl}{z}\right) + ...  \right]
\label{a_2}
\eeq
 is the dressed backscattering amplitude and $Z =1/(1+\lambda)$  is the quasiparticle spectral weight.
However, if we go beyond logarithmic accuracy, we find that $\Sigma^{(3)}_{\omega^2,\text{b}}(\omega_m)$ contains contributions that come from $|\q| \sim M$ and cannot be expressed via the  backscattering vertex. Only very close to a QCP, where $M$ is vanishingly small, these non-logarithmic terms can be absorbed into the dressed backscattering amplitude.

 For the diagram of Fig.~\ref{planar3loopvertices}(a), which is a candidate for the forward-scattering contribution,  the dressed vertex (the blue box) vanishes at external $q, q_2=0$, and to reproduce $\Sigma^{(3)}_{\omega^2,\text{a}}(\omega_m)$ one has to keep $|\q_2| \sim M$. This implies that $\Sigma^{(3)}_{\omega^2,\text{a}}(\omega_m)$ does not come from forward scattering.  This again holds at a finite $M$.   The vanishing of the dressed vertex  at $q, q_2 =0$ also explains why the three-loop self-energy  $\Sigma^{(3)}_{\omega^2,\text{a}}(\omega_m)$ does not contain a logarithm.

We therefore see that in a Fermi liquid, only planar diagrams survive in the limit $N \to \infty$, but only a portion of the self-energy  from these diagrams comes from forward scattering and backscattering. The converse is also true: one can easily verify that the self-energy $\Sigma^{(3)}_\text{a}(\omega_m)$ from the non-planar diagram in Fig.~\ref{planar3loopvertices}(c) does come from backscattering. However, this vertex renormalization is small in $1/N$.

 Below we keep  $M$ finite and compute the full self-energy to all orders in the interaction to logarithmic accuracy.
 To this accuracy, we can neglect renormalizations from ``near-forward-scattering"  and  ``near-back-scattering"  processes and compute the renormalization of the backscattering amplitude keeping only the largest power of $\ln{(\omega_m)}$ at each order in the loop expansion. The fact that the self-energy can be obtained from the renormalized backscattering amplitude (to the logarithmic accuracy) again implies that the problem is effectively one-dimensional: the relevant internal fermions have momenta $\pm \k$ up to small deviations of order $\omega$, i.e., they only move along a single direction. 
 We note that for this calculation we do not need to set $N \to \infty$ as the $1/N$ terms are outside the logarithmic accuracy.

\section{The full result for the self-energy to all orders in the interaction}
\label{flselfsecrep}

\begin{figure*}
\centering
\includegraphics[width=.8\textwidth]{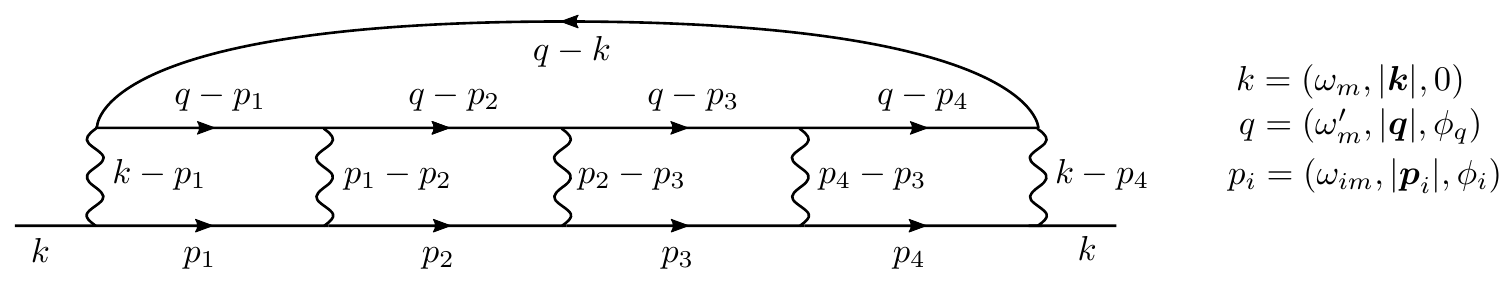}
\caption{Diagram contributing to $\Sigma(\omega)$ with 4 Cooper bubbles. Labels indicate four-momenta, interaction lines are RPA dressed. }
\label{maindiagram}
\end{figure*}

Let us now compute the fully renormalized backscattering contribution to the self-energy by extracting the most logarithmically singular contributions to $\Sigma_{\omega^2, \text{bs}}$ at each loop order. One can straightforwardly verify that these contributions arise from the diagrams shown in Fig.\ \ref{maindiagram}, which particle-particle (Cooper) ``bubbles".  We label  them as  ``Cooper diagrams''.
An example of a  Cooper diagram is shown in Fig.\ \ref{maindiagram}. A generic  diagram with
$n$ Cooper bubbles contains $\ln(\omfl/|\omega_m|)^n$. Our goal is to sum up this series.  We will see that the series
is not geometrical, because the interaction is momentum-dependent.

We proceed with the evaluation of the Cooper diagrams. An inspection of these diagrams %Dima shows% 
yields the following general strategy for the calculation of the $O(\omega^2_m)$ term in the self-energy to logarithmic accuracy (see App.\ \ref{maincalcdetailssec} for details):
    we select one cross-section, from  which we take the Landau-damping term (this gives the overall $\omega_m^2$ in $\Sigma_{\omega^2, \text{bs}}$) and sum up series of Cooper logarithms on both sides of this cross-section, %Dima rewritten since the \Gamma_bs^2 was not really discussed yet% 
which produces a factor $\Gamma^2_{\text{bs}}$. For the calculation of the Landau damping in the selected cross-section with internal momentum $\p_i$ in the Cooper bubble, we set  ${\q}$  in Fig.\ \ref{maindiagram} nearly transverse to the external ${\k}$ and $\p_i$ nearly parallel to ${\k}$. For the other Cooper bubbles, we only use a static interaction between fermions on the Fermi surface, 
 \bea
\label{interactionform}
D(p_1 - p_2) \simeq  -g k^2_F \left[\epsilon^2 +  4 \sin^2\!\left(\frac{\phi_{1} - \phi_{2}}{2}\right) \right] \ , 
\eea
where $\phi_i$ are the angles of $\p_i$ measured relative to $\k$. We  evaluate each particle-particle bubble $\Pi_\text{pp}$  to logarithmic accuracy and for $v_F q \gg \omega_m'$, where $\omega'_m$ is the frequency component of $q$ in Fig.\ \ref{maindiagram}. 
Under these conditions, the integration of the two fermionic propagators  in the particle-particle bubble with $q-p_j$ and $p_j$ over internal frequency $\omega_{jm}$ and over $|\p_j|$  yields (see App.\ \ref{bubblesapp})
%Dima: made into equation for easier reading%
\begin{align}
\label{Pippmain}
 \Pi_{\text{pp}} (|\q|, \phi_j) = Z \rho\ln\left(\frac{\omfl}{v_F |\q| |\sin{\phi_j}| }\right) \ . 
 \end{align} 
 We assume and verify that typical values of {\it all} $\phi_j$ are of order $\epsilon$.
 To logarithmic accuracy, we can then approximate $\Pi_{\text{pp}} (|\q|, \phi_j) \simeq \Pi_{\text{pp}} (z) = Z \rho
 \ln(\omfl/z)$, where $z =v_F |\q|\epsilon$. Then the remaining integrations over $\phi_{j}$ in the Cooper ladder 
 involve only the interactions.   This leads to the following expression for the backscattering amplitude, with $L(z) = \ln(\omfl/z)$: 
 \begin{widetext}
\begin{align}
\label{Tfirst}
&
\Gamma_{\text{bs}} (z) \equiv - \frac{g}{k_F^2} \sum_{n = 0}^\infty \left(\frac{g\rho Z L(z)}{k^2_F}\right)^n
  \cdot \frac{1}{(2\pi)^n} \int_0^{2\pi} d\phi_1 d\phi_2 \hdots d\phi_n\frac{1}{\epsilon^2 + 4 \sin^2\left(\frac{\phi_1}{2}\right)}  \times \frac{1}{\epsilon^2 + 4 \sin^2\left(\frac{\phi_2 - \phi_1}{2}\right)} \times \hdots  \\ & \notag  \times \frac{1}{\epsilon^2 + 4\sin^2\left(\frac{\phi_n - \phi_{n-1}}{2}\right)} \times \frac{1}{\epsilon^2 + 4 \sin^2\left(\frac{\phi_{n}}{2}\right)} \  .
\end{align}
To leading order in $\epsilon$, the integrand  can be approximated as 
\begin{align}
\label{phiint}
&\frac{1}{(2\pi)^n \epsilon^{n+1}} \int_{-\infty}^\infty d\phi_1 \hdots d\phi_n \frac{\epsilon}{\epsilon^2+ \phi_1^2} \times \frac{\epsilon}{\epsilon^2+(\phi_1 - \phi_2)^2} \times \frac{\epsilon}{\epsilon^2+ (\phi_2 - \phi_3)^2} \times \hdots \times \frac{\epsilon}{\epsilon^2 + (\phi_{n-1} - \phi_n)^2} \times  \frac{\epsilon}{\epsilon^2+\phi_n^2} \notag  \\ &  = \frac{1}{(2\pi)^n \epsilon^{n+2}} \times \frac{\pi^n}{(n+1)} 
%Dima added factor 2
= \frac{1}{n+1} \frac{1}{(2\epsilon)^{n}} \times \frac{k^2_F}{M^2} \ .
\end{align}
\end{widetext}
The integrals in Eq.\ \eqref{phiint} were solved by going to Fourier space, where the convolutions of Lorentzians become products of exponentials. The expression for the backscattering amplitude then reduces to  
\begin{align}
\label{Tsumfirst}
\Gamma_{\text{bs}}(z)
 \simeq - \frac{g}{M^2} \sum_{n = 0}^\infty \left(
\tilde{\lambda} 
  L(z)\right)^n  \frac{1}{n+1} , \quad  \tilde\lambda \equiv  \lambda Z = \frac{\lambda}{1+ \lambda}\  .
\end{align}
We see that
 $\lamt$ becomes the effective coupling controlling the perturbation theory. 
 As long as $\lamt L(z) < 1$, the summation in Eq.\  \eqref{Tsumfirst} converges, and we obtain
\begin{align}
\label{Tlognormal} \Gamma_{\text{bs}}(z) =  \frac{g}{M^2}\left(\frac{\ln\left(1- \lamt L(z)\right)}{\lamt L(z)} \right) \ .
\end{align}
The full self-energy from backscattering then becomes
\begin{align}
\label{Sigmap2}
&\Sigma_{\omega^2, \text{bs}}(\omega_m) = i \sign(\omega_m) \frac{\lambda}{\pi} \frac{\omega^2_m}{\omfl}   \int_{\omega_m}^{\omfl}  \frac{dz}{z}
 \left[\frac{M^2}{g} \Gamma_{\text{bs}} (z)\right]^2  \notag \\
 & =   i \sign(\omega_m) \frac{\lambda}{\pi}  \frac{\omega^2_m}{\omfl}
 \int_{\omega_m}^{\omfl}  \frac{dz}{z} \left(\frac{\ln\left(1- \lamt L(z)\right)}{\lamt L(z)}\right)^2  \ .
\end{align}
At $\lamt \rightarrow 0$, we get 
\begin{align}
\label{wcresult}
\Sigmab(\omega_m) = i\sign(\omega_m) \frac{\lambda}{\pi} \frac{\omega^2_m}{\omfl} \ln\left(\frac{\omfl}{|\omega_m|}\right) \ .
\end{align}
This is exactly a half of the one-loop result, which is expected, as Eq.\ \eqref{Sigmap2} only contains the backscattering contribution. {The weak-coupling result \eqref{wcresult} is only valid for $\lambda \ln(\omfl/|\omega_m|) \ll 1$; for small frequencies, $\lambda \ln(\omfl/|\omega_m|)$ becomes $O(1)$, and one has to solve the full integral \eqref{Sigmap2} instead. }

\subsection{Repulsive pairing interaction}
\label{repulsive}

\begin{figure}
\centering
\includegraphics[width=\columnwidth]{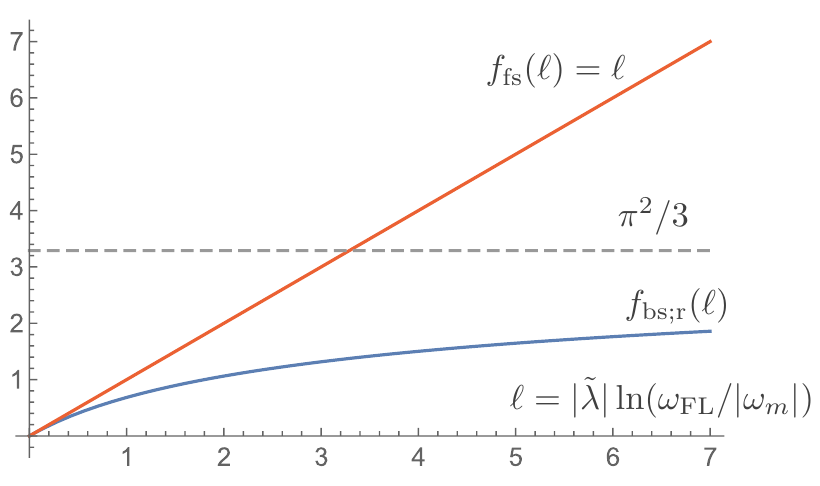}\caption{The functions
$f_\text{bs;r}(\ell)$ and $f_\text{fs}(\ell)$ 
 for the toy model with repulsive pairing interaction.  }
\label{frepulsiveplot}
\end{figure}

We first analyze Eq.~(\ref{Sigmap2}) for a toy model, in which we assume that ${\tilde \lambda}$ is negative, i.e., that the pairing interaction is repulsive.  In this case,  the series in Eq.\ (\ref{Tsumfirst}) converges for all $z$, even the smallest ones. The integral over $z$ in Eq.~\eqref{Sigmap2} can then be evaluated exactly, and the result is
\begin{align}
\label{Sigmawithxrep}
\Sigmab(\omega_m) &= i \sign(\omega_m) 
 \frac{\omega^2_m }{ Z \pi \omfl} f_\text{bs;r}(\ell) \ ,
 \end{align}
  where
  \beq
   \ell = 
   |\lamt| \ln(\omfl/|\omega_m|)
\eeq
 and
 \beq
f_\text{bs;r}(\ell) =
- \frac{(1+\ell) \ln^2(1+\ell)}{\ell} - 2\text{Li}_2(-\ell) \ ,
\eeq
with  $\text{Li}_2$ the Polylogarithm.
The forward scattering contribution, in the same units, is
\begin{align}
\Sigma_{\text{fw}}(\omega_m) = i\text{sign}(\omega_m) 
 \frac{\omega^2_m}{Z \pi \omfl}  f_\text{fs}(\ell), \quad f_{\text{fs}}(\ell) = \ell \ .
\end{align}
A plot of $f_\text{bs;r}, f_\text{fs}$ is shown in Fig.\ \ref{frepulsiveplot}. For small $\ell$, i.e., fairly large frequencies, $f_\text{bs;r} = f_{\text{fs}}$ as expected. For smaller frequencies (increasing $\ell$), $f_\text{bs;r}(\ell)$ does not grow linearly, but is bounded, and can be approximated by the limiting form
\begin{align}
f_\text{bs;r}(\ell) = \frac{\pi^2}{3} - \frac{\ln^2(\ell)}{\ell} + \mathcal{O} \left( \frac{\ln(\ell)}{\ell}
\right) \ .
\end{align}
Therefore, for a repulsive pairing interaction, the backscattering rate incures a logarithmic suppression compared to forward scattering; at large $\ell$ the full rate is reduced  to the forward scattering part, i.e., by a half compared to the one-loop result.

 Upon analytical continuation to the real axis, $i\omega_m \to \omega + i\delta$, Eq.~(\ref{Sigmawithxrep}) becomes the expression for the scattering rate  at a low frequency 
 \begin{align}
\label{Imsigmarep}
{\text{Im}} \left[\Sigma_{\text{bs;r}}^R (\omega)\right] &= - 
\frac{\omega^2}{Z\pi \omfl} f_\text{
bs;r}(\ell) \ ,
\end{align}
 where now $\ell = |{\tilde \lambda}| \ln\left(\omfl/|\omega|\right)$.

\subsection{Attractive pairing interaction}
\label{attractivenormalstatesec}
\begin{figure}
\centering
\includegraphics[width=\columnwidth]{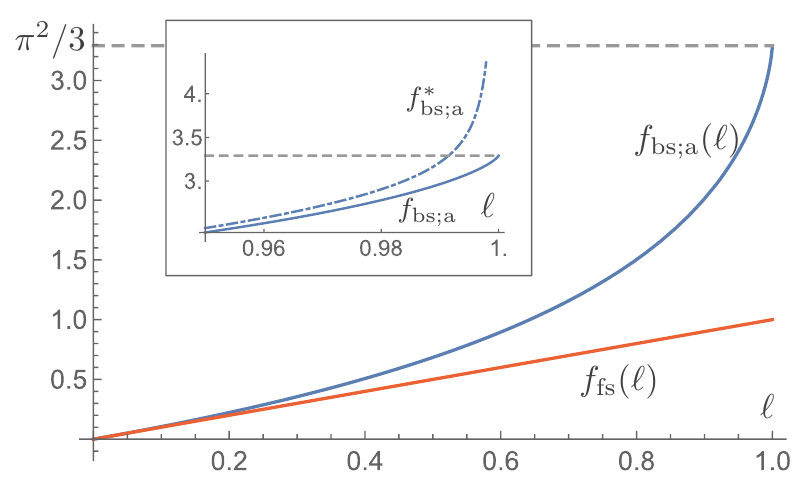}
\caption{
 The function  $f_{\text{bs;a}}$ to leading order in $\epsilon$,  from Eq. (\ref{sigmaattleading}). 
  $\ell = 1$ corresponds to $\omega = \Delta_0$. The inset shows the modification of 
  $f_\text{bs;a}$ due to extra contributions beyond the leading order in $\epsilon$,  
   see  Eq.~\eqref{fbsstarall}.
}
\label{fattractiveplot}
\end{figure}

Let us now consider the physical case of attractive pairing interaction, $\lamt >0$. Now we have
\begin{align}
\label{sigmaattleading}
\Sigmab &= i\text{sign}(\omega_m) \frac{\omega^2_m}{Z\pi \omfl}
f_\text{bs;a}(\ell) \ , \nonumber \\
f_\text{bs;a}(\ell) &= \frac{ (\ell - 1)\ln^2(1-\ell)}{\ell} + 2\text{Li}_{2}(\ell) \ ,
\end{align}
and  $f_{\text{bs;a}}(\ell)$ is well-defined  only for $\ell \leq 1$.
At larger $\ell$ the results must be modified by including superconductivity, as we show below. A plot of $f_\text{bs;a}(\ell)$ is presented in Fig.\ \ref{fattractiveplot}. Again, for small $\ell$, 
$f_\text{bs;a} \simeq \ell$, but at
$\ell \lesssim 1$ it grows faster.  In the limit $\ell \rightarrow 1$,
\begin{align}
f_\text{bs;a}(\ell) = \frac{\pi^2}{3} - (1 - \ell) \ln^2(1- \ell) + \mathcal{O}\left(\ln(1-\ell) (1- \ell)\right) .
\end{align}
$
f_\text{bs;a}(1) = \pi^2/3$ is finite, but the slope $df_\text{bs;a}(\ell)/d\ell$  is logarithmically divergent at
  $\ell \rightarrow 1$. As a curiosity, we note that $f_{\text{bs;r}}$ and $f_{\text{bs;a}}$ are related as
\beq
\label{fssym}
f_{\text{bs;r}}(\ell) =f_\text{bs;a}\left(\frac{\ell}{1+\ell} \right) \ .
\eeq
This can be derived by a simple substitution in the integral defining $f_\text{bs;a}$. 

Let us look at the behavior of
$f_{\text{bs;a}} (\ell)$  as
%Dima still \ell<= 1$  $\ell \geq 1$ 
 $ \ell \rightarrow 1$ more carefully. We recall that
$f_{\text{bs;a}} (\ell)$   has been obtained using the expression for the backscattering amplitude $\Gamma_{\text{bs}} (z)$, Eq.~\eqref{Tlognormal},  valid to the leading order in $\epsilon = M/k_F$. We now %Dima verify % 
check whether the dependence of $f_{\text{bs;a}} (\ell)$ on $1-\ell$ gets modified if we compute $\Gamma_{\text{bs}} (z)$ beyond  $\mathcal{O}(\epsilon)$.  The reasoning for doing this is the following:  to leading order in $\epsilon$, partial amplitudes in pairing channels with different angular momentum are all equal, and this gives rise to the appearance of 
  the factor $1/(n+1)$ in the series for $\Gamma_{\text{bs}} (z)$ in \eqref{Tlognormal}.
 Beyond the order $\mathcal{O}(\epsilon)$, partial pairing amplitudes do differ, and the largest one is in the $s$-wave channel.
 If we would only include this channel and ignore all other pairing components, we would find that the logarithmical series 
 becomes geometrical and $f_{\text{bs;a}} (\ell) \propto 1/(1-\ell)$ (see Eq.~(\ref{sigmaattsubleading_1}) below).  For a small but finite $\epsilon$, we therefore expect a crossover from 
  $f_{\text{bs;a}} (\ell)= \pi^2/3- (1-\ell) \ln^2(1- \ell) +...$ at $1-\ell\leq 1$ to  $
   f_{\text{bs;a}} (\ell) \propto 1/(1-\ell)$ at the smallest $1- \ell$. To describe this crossover,  we now evaluate $\Gamma_{\text{bs}}(z)$ in next-to-leading order in $\epsilon$.
  This can be achieved by angular momentum decomposition. 
  Namely, %Dima: V was not used before, thus inserted this here%
  the effective interaction on the Fermi surface, 
\begin{align}
V(\phi) &= 
\lamt \frac{2\epsilon}{\epsilon^2 + 4 \sin^2(\phi/2)}\ , 
\end{align}
 can be decomposed into angular momenta as
\begin{align}
\label{Veffangular}
V_m = \int_0^{2\pi} \frac{d\phi}{2\pi} V(\phi) \exp(-i\phi m) \ .
\end{align}
 To a very good approximation (see  App.\ \ref{Tmatrixapp})
 \begin{align}
\label{Veffangular_1}
V_m  \approx \lamt \exp(- |m| \epsilon) \ .
\end{align}
At $\epsilon \rightarrow 0$, all components  $V_m$ become degenerate. However, for a nonzero $\epsilon$,
   this is not quite the case yet, and  the $s$-wave  component 
     $V_0 \simeq \lamt$ is largest.
A straightforward analysis then shows that the series for
 $\Gamma_{\text{bs}}(z)$ can be rewritten as
\begin{align}
\notag
 \Gamma_{\text{bs}}(z) &= - \frac{1}{\rho} \sum_m V_m \sum_{n = 0}^\infty \left[ V_m L(z)\right]^{n} \\ &= - \frac{1}{\rho} \frac{V_0}{1-V_0 L(z)} - \frac{2}{\rho} \sum_{m = 1}^\infty \frac{V_m}{1- V_m L(z)},
\label{remsum}
\end{align}
where in the second line we singled out the $s$-wave part. The remaining sum in Eq.\ \eqref{remsum} can be evaluated using the Euler-Maclaurin formula, and  the result is
\begin{align}
\label{Tnextto}
&\Gamma_{\text{bs}}(z) \simeq    \\ & \notag -\frac{g}{M^2}
\left[-\frac{\ln\left( 1- \lamt L(z)
 +
  \epsilon
 \right)}{\lamt L(z)} + \frac{\epsilon}{2} \frac{1}{1- \lamt L(z)} \right] .
\end{align}
Eq.\ \eqref{Tnextto} shows that the $\ln\left( 1- \lamt L(z)\right)$ dependence of the
backscattering amplitude is the consequence of the near degeneracy of all
partial pairing amplitudes. Once this degeneracy is lifted at a finite $\epsilon$,
  the backscattering amplitude develops a conventional $s$-wave pole,
 albeit with a small weight $\epsilon/2$. Inserting this
  $\Gamma_{\text{bs}} (z)$  into the integral for  $\Sigmab$, Eq.~(\ref{Sigmap2}), we obtain, 
 \begin{align}
\label{sigmaattsubleading_1}
&\Sigmab = i \sign(\omega_m) \frac{\omega_m^2}{Z \pi \omfl}
f^*_\text{bs;a}(\ell),  
\end{align} 
where \begin{align}
\label{fbsstarall}
f^*_\text{bs;a}(\ell) \simeq f_\text{bs;a}(\ell) + \frac{\epsilon^2}{4(1-\ell)} \ .
\end{align} 
 At $1-\ell \lesssim \epsilon$, 
 \begin{align}  \notag 
f^*_{\text{bs;a}}(\ell) \simeq   &\frac{\epsilon^2}{4(1- \ell)}  + \frac{\pi^2}{3}  + {\epsilon} \ln{(1-\ell + \epsilon)}\ln{(1-\ell)}
 \\ & - (1-\ell +\epsilon) \ln^2 (1-\ell + \epsilon) \ .
 \label{a_5}
\end{align}
$f_\text{bs;a}(\ell)$ and $f^*_\text{bs;a}(\ell)$ are plotted in the inset of Fig.\ \ref{fattractiveplot}. 
We see that the self-energy diverges as $1/(1-\ell)$  at $\ell \to 1$, as is expected for an $s$-wave superconductor
~\cite{PhysRevB.42.10211,Larkin2008}, but in our case the divergent term appears with the prefactor $\epsilon^2$ \footnote{We note that the prefactor of the $s$-wave term gets renormalized ($N \rightarrow N-1$ in the expression for $\omfl$) by ladder diagrams without a loop, in which a line, expressing an outgoing fermion, is attached to the upper part of the Cooper ladder \cite{haussmann1993crossover}. Such a diagram has the same number of logarithms, but involves a large momentum transfer $2k_F$ and therefore renormalizes the $s$-wave part only}.

Like before, upon analytical continuation to the real axis, $i\omega_m \to \omega + i\delta$, Eq.\ (\ref{sigmaattsubleading_1}) becomes the expression for the backscattering contribution to
Im[$\Sigma^R(\omega)$]:
\beq
\text{Im}\left[\Sigma_{\text{bs;a}}^R(\omega)\right] =
- \frac{\omega^2}{Z \pi \omfl} 
  f^*_\text{
  bs;a}(\ell) \ , 
\label{sigmaattsubleading_2}
\eeq
where on the real frequency axis $\ell = {\tilde \lambda} \ln({\omfl/|\omega|})$.  Eq.~(\ref{sigmaattsubleading_2}) shows that the scattering rate increases when $\omega$ decreases ($\ell$ increases)  and
becomes singular at $\ell =1$.

The  enhancement of $\text{Im}[\Sigma_{\text{bs;a}}^R(\omega)]$ near $\ell =1$
can be interpreted as follows:
to the lowest order in the interaction, the rate comes from the
 decay of a particle into two particles and one hole; $\text{Im}[\Sigma_{\text{bs;a}}^R(\omega)]$ measures the phase space for such a process. By energy conservation, the three outgoing states must  have energies smaller than $\omega$.
When Cooper pair formation sets in, the two electrons in the particle-particle ladder form a tightly bound Cooper pair, and there are only two decay products left. Therefore, the phase space restriction is less severe for small $\omega$,   and the scattering rate increases.

Before concluding this section, we discuss the range of validity of Eqs.~(\ref{sigmaattsubleading_1}) and (\ref{sigmaattsubleading_2}).  The expression for the self-energy has been obtained with  logarithmic accuracy,
 i.e., at each loop order we neglected terms with the same power of ${\tilde \lambda}$, but a smaller power of $\ln(\omfl/|\omega|)$. This is valid when ${\tilde \lambda} \ll 1$ and $|\omega| \ll \omfl$.
The condition on $\omega$ does not pose a restriction as throughout the paper we assume that $\omfl$ is finite and focus on the self-energy at the smallest $\omega$.  The condition ${\tilde \lambda} = \lambda/(1+\lambda) \ll 1$ implies  weak coupling, $\lambda \ll 1$.  Because $\lambda = g/(4\pi v_F M) \sim g/(E_F \epsilon)$, we need to keep $g/E_F$ small to satisfy both $\lambda \ll 1$ and $\epsilon \ll 1$. 
  Weak coupling is advantageous to us because,  for ${\tilde \lambda} \ll 1$, 
there is a wide range of frequencies where on the one hand $|\omega| \ll \omfl$ and on the other  $\ell  = {\tilde \lambda}  \ln(\omfl/|\omega|) <1$,  hence Eqs. (\ref{sigmaattsubleading_1}) and (\ref{sigmaattsubleading_2}) are applicable. This is indeed the consequence of the fact that superconductivity at weak coupling emerges at an exponentially small $\omega \sim \Delta_0 =\omfl \exp(-1/{\tilde \lambda})$.   We emphasize that this only holds in a Fermi liquid regime, at small enough $g/E_F$. Very close to a QCP, $\lambda$ necessarily becomes
 large,
 and the whole Fermi liquid range $\omega < \omfl$ falls into $\omega < \omega_{\text{IN}} \sim \lambda^3 \omfl$.  The pairing instability in this case emerges at a 
   larger $\omega \leq \omega_{\text{IN}}$ (Refs. \cite{PhysRevLett.77.3009,Moon2010,PhysRevB.91.115111,CHUBUKOV2020168142}), hence there is no range of applicability for our analysis.

\section{Self-energy in the superconducting state}
\label{scscatesec}

We now show that the singularity of the self-energy at $\ell=1$ gets regularized once we take into consideration the fact that the ground state is actually an $s$-wave superconductor.
 In practical terms, superconductivity implies that for the calculation of the backscattering amplitude and the fermionic self-energy one has to use normal and anomalous Green's functions
\begin{align}
\label{Gmeanfieldmain}
G(\omega_m, \k) &= -
 \frac{i Z^{-1}\omega_m + \xi_\k}{Z^{-2}\omega^2_m + \xi_\k^2 + \Delta_0^2}\ ,  \\ \notag    F(\omega_m, \k) &=  \frac{\Delta_0}{Z^{-2} \omega^2_m + \xi_\k^2 + \Delta_0^2} \ .
  \end{align}

The  three key processes that determine the decay of a quasiparticle in a superconductor are 
(i)
 scattering by amplitude and phase fluctuations of the  superconducting order parameter,  
 (ii) scattering by a (potential) resonance mode in the nematic $D(\omega, {\q})$ at (real) $\omega = \omega_{\text{res}} <3\Delta_0$  and (iii) scattering due to a decay into particles and holes (this process starts at $\omega =3\Delta_0$).
 For a charged superconductor, scattering by phase fluctuations is affected by long-range Coulomb interaction, which converts a Goldstone phase fluctuation mode into a plasmon (still gapless in 2D).  However, we neglected the effects of long-range Coulomb interaction in the calculations assuming a normal state at $T=0$, and for consistency we also have to neglect this interaction in a superconductor.
   
In general,  all three processes are important at $\omega_m \gtrsim \Delta_0$,  and the full-fledged
 evaluation of $\Sigma(\omega_m \gtrsim \Delta_0)$
 is a complicated endeavor. 
Since our primary goal is to understand how superconductivity affects 
 singular behavior of the self-energy that we found in the previous section, we limit ourselves to the scattering processes which cut off the singularity.

 We will directly focus on the imaginary part of the backscattering self-energy  in real frequencies, 
  $\text{Im}[\Sigma_\text{bs;a}^R(\omega)]$.  Consider first the contribution to  $\text{Im}[\Sigma_\text{bs;a}^R(\omega)]$ arising from the sum of all angular momentum channels,
  Eq.~\eqref{sigmaattleading}. 
  This expression shows a slope singularity
   at $\ell =1$, however,
     since it was obtained with the logarithmic accuracy, it is only valid for $ \ell 
     \lesssim 1- \lamt$.  At the boundary of this regime, the self-energy \eqref{sigmaattleading} is of the same order as
the one-loop result, Eq.\ \eqref{a_1}. Therefore, to estimate this part of the self-energy
  at $1-\ell <\lamt$,  it is sufficient to re-evaluate the one-loop diagram of Fig.\ \ref{diagtable}(b) in the superconducting state.
     We find (see App.\ \ref{Imsigma1app} for details)
\begin{align}\label{Imsigma3delta}
&-\text{Im}\left[\Sigma_\text{bs;a}^{R}(\omega)\right] \sim  \\ &   \frac{\Delta_0^2} {\omfl} \lamt \ln\left( \frac{\omfl}{\Delta_0}\right) \times \sqrt{ \frac{\omega - 3\Delta_0}{ \Delta_0}} \theta(\omega - 3\Delta_0) \ .  \notag
\end{align}  
We have suppressed the dependence on $Z \simeq 1$. The scattering rate of Eq.\ \eqref{Imsigma3delta} starts at $\omega = 3\Delta_0$, which is transparent as the three scattering products, two quasiparticles and a quasihole, all have an energy gap $\Delta_0$. At $\omega - 3\Delta_0 \simeq \Delta_0$, i.e., at  $1-\ell \sim \lamt$,  Eq.\ \eqref{Imsigma3delta} becomes of the same order as the normal state result, Eq.~\eqref{sigmaattleading}.  
We see therefore that the portion of the self-energy that comes from all scattering channels smoothly evolves from
 Eq.~\eqref{sigmaattleading} to Eq.~(\ref{Imsigma3delta}).   We also computed the contribution to the self-energy 
  from the resonance in $D(\omega, \q)$ in the superconducting state.  The resonance develops at $\omega_{\text{res}} = \Delta_0 \times \left[1 + \mathcal{O}(\epsilon k_F/(N \rho g)^{1/2})\right]$.   Adding this contribution, we find that $\text{Im}[\Sigma_{\text{bs;a}}^{R}(\omega)]$ extends down to $\omega_{\text{res}}$, but near $\omega \sim 3\Delta_0$ it is given by 
   Eq.\ (\ref{Imsigma3delta}) up to small corrections.   The form of  $\text{Im}[\Sigma_{\text{bs;a}}^{R}(\omega)]$ at $\omega_{\text{res}} \leq \omega \leq 3\Delta_0$ is rather involved and we will discuss it in a separate report. Ultimately, the complicated form of the self-energy is tied to the abundance of low-energy collective excitations in a superconductor with long-ranged interactions, as also recognized in Refs.\ \cite{PhysRevB.62.11778, KleinNpj2019}. 
 
To find the correct cutoff of the $s$-wave pole, Eq.\ \eqref{sigmaattsubleading_1}, evaluation of the one-loop diagram is insufficient, and one 
 needs to re-evaluate the  full backscattering amplitude $\Gamma_{\text{bs}}$ using  $\Pi_\text{pp}= GG + FF$. 
    The key new effect coming from this calculation is the scattering of a quasiparticle 
    by  massless (Goldstone) phase fluctuations of the superconducting order parameter. Evaluating the $s$-wave component of $\Gamma_{\text{bs}}$ and substituting the result into the self-energy, we obtain
\begin{align}
\label{Imsigmas}
-
 \text{Im}\left[\Sigma_\text{bs;a}^{R}(\omega)\right] \sim 
  \frac{\epsilon^2}{\lamt} \frac{\Delta_0^2}{\omfl}  \times \theta(\omega - \Delta_0)\  .
\end{align}
This contribution to the self-energy starts at $\omega = \Delta_0$, since the phase mode is gapless. 
  For  $1- \ell \simeq \lamt$, i.e., $\omega \gtrsim \Delta_0$, this self-energy 
matches  with the $s$-wave piece  $\omega^2/\omfl \times \epsilon^2/(1- \ell)$, obtained in the normal state calculation,  Eq.\ \eqref{sigmaattsubleading_1}.
  A similar expression for the self-energy at $T=0$  in a 3D superfluid has recently been computed in Ref.~\cite{PhysRevLett.124.073404}, building on older work of Ref.\ \cite{haussmann1993crossover}. The contribution from the self-energy to the density of states of an $s$-wave superconductor has been obtained in Refs.
  \ \cite{PhysRevB.42.10211, Larkin2008}.

Collecting  the results, we find the  full self-energy from backscattering in the form 
\begin{widetext}
\begin{align} \label{ImsigmafinalresultN}
&0< \  \ell \lesssim 1- \mathcal{O}(\lamt): \quad &&\text{Im}\left[\Sigma_{\text{bs;a}}^R(\omega)\right] =
-\frac{\omega^2}{\pi \omfl}   f^*_\text{bs;a}(\ell),  \quad
 f^*_\text{bs;a}(\ell) \simeq  \frac{ (\ell - 1)\ln^2(1-\ell)}{\ell} + 2\text{Li}_{2}(\ell) + \frac{\epsilon^2}{4(1-\ell)}, \\ &
1- \mathcal{O}(\lamt)\lesssim \  \ell < 1: \quad &&\text{Im}\left[\Sigma_{\text{bs;a}}^R(\omega)\right] =  -  \frac{\Delta_0^2} {\omfl} \lamt \ln\left( \frac{\omfl}{\Delta_0}\right) \times \sqrt{ \frac{\omega - 3\Delta_0}{ \Delta_0}} \theta(\omega - 3\Delta_0) -   \frac{\epsilon^2}{\lamt} \frac{\Delta_0^2}{\omfl}  \times \theta(\omega - \Delta_0) .
\label{ImsigmafinalresultS}
\end{align}
\end{widetext}

\begin{figure*}
\centering
\includegraphics[width=.95\textwidth]{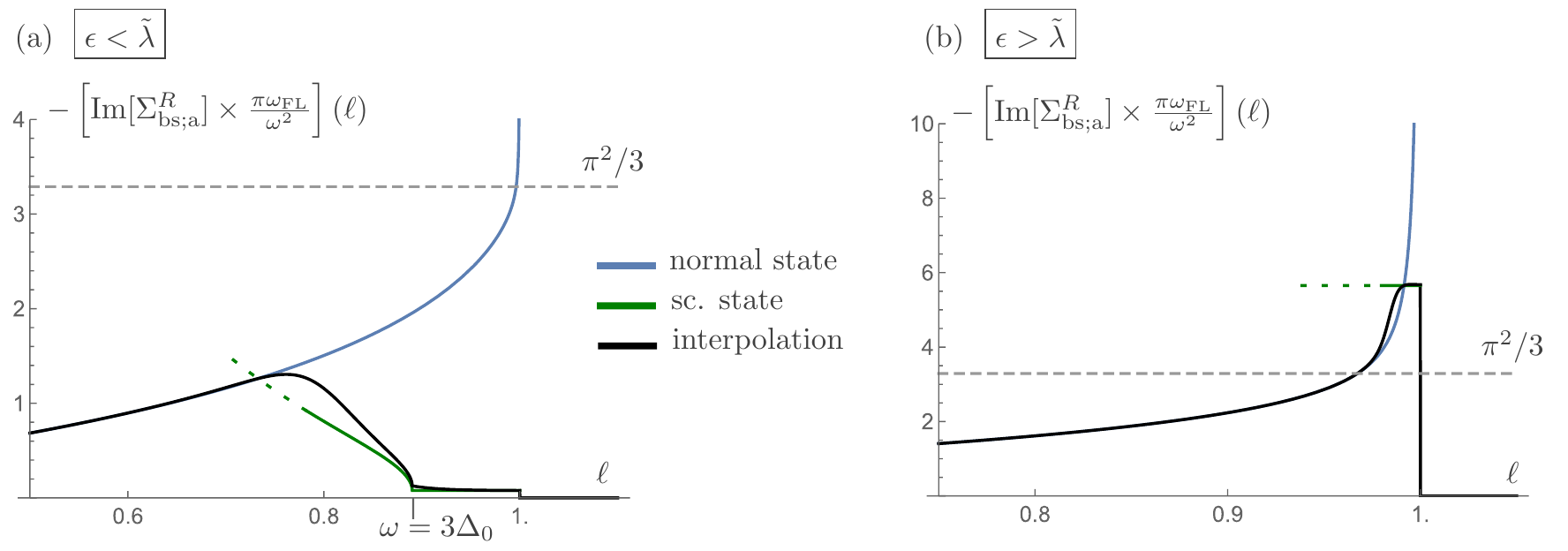}\caption{Backscattering self-energy divided by $\omega^2$ as a function of the logarithmic variable $\ell$. Blue curves are the normal state result (Eq.\ \eqref{ImsigmafinalresultN} and Fig.\ \ref{fattractiveplot}), green curves correspond to the superconducting result, Eq.\ \eqref{ImsigmafinalresultS}.  The black line is a possible interpolation representing the correct physical backscattering rate. (a) Sum of angular momentum channels dominates, used parameters: $\epsilon = 0.05,  \lamt = 0.1$. (b): $s$-wave channel dominates, used parameters: $\epsilon = 0.3, \lamt = 0.05$.  }
\label{interpolfig}
\end{figure*}

We sketch $[\Sigma_{\text{bs;a}}^R(\omega)]/\omega^2$ in Fig.\ \ref{interpolfig}. In panel (a) we set $\epsilon < \lamt$.  In this situation,  the contribution from all angular channels dominates over the $s$-wave part  for all $\ell$. In panel (b) we set  $\epsilon > \sqrt{\lamt}$. Here the contribution from the $s$-wave channel is the dominant one. We see that in this situation   Im$[\Sigma_{\text{bs;a}}^R(\omega)]/\omega^2$ has a sharp peak at $\ell \leq 1$. This can be directly verified in ARPES experiments.

Note that while  
%While the
 the imaginary part of the self-energy in real frequencies
   vanishes at $|\omega| < \Delta_0$, the Matsubara self-energy (which can be obtained from Quantum Monte Carlo) is non-zero.
     At $\omega_m \ll \Delta_0$, $\Sigma_{\text{bs;a}}(\omega_m)$ scales as 
     %will be an odd function of
      $\omega_m/\Delta_0$. At larger $\omega_m \sim \Delta_0$, $\Sigma_{\text{bs;a}}(\omega_m)$ passes through a maximum.
      %at the smallest frequencies, but still inherit the maximum at $\omega_m \simeq \Delta_0$   from $\text{Im}\left[\Sigma_{\text{bs}}^R(\omega)\right]$. (\mr{more detail needed?)}.

\section{Conclusions}
\label{concsec}

In this work we computed the fermionic self-energy in a Fermi liquid near a nematic QCP.  We considered  the 
 model of fermions interacting with soft fluctuations of a nematic order parameter, and extended it to $N$ fermionic flavors. The leading contributions to the self-energy at $N \to \infty$ come from planar diagrams, as was determined before at a QCP. In a Fermi liquid, the contributions from planar diagrams contain series of logarithms. We identified the 
leading logarithmic  contributions at all orders as coming from the fully dressed backscattering amplitude, i.e., 
they describe one-dimensional scattering  processes. We further argued that the subleading contributions (the terms with a smaller power of $\ln(\omega)$ at each loop order) come from scattering with a finite momentum transfer counted from either backscattering or forward scattering.
 
We found that the logarithmic series are non-geometrical and summed them up using several computational tricks. 
We first considered  a toy model with repulsive pairing interaction, and then a realistic model with attractive pairing interaction. For the latter case we found two contributions to the self-energy%Dima,%
: one is the combined contribution from multiple pairing channels with near-equal attraction, and the second is the special contribution from an $s$-wave pairing channel, where the pairing interaction is a bit larger than in other channels. We first computed the two terms assuming the normal state and found that perturbation theory holds only above a certain frequency, comparable to superconducting gap $\Delta_0$.  At the edge, both contributions become non-analytic and the one from $s$-wave channel diverges. We then extended the analysis to include superconductivity at $T=0$ and showed that the  would-be divergencies get regularized. In particular, the  divergence in the  $s$-wave channel is  regularized by scattering of a fermion by phase fluctuations of the superconducting order parameter. We obtained the full expression for the imaginary part of the retarded self-energy in real frequencies (the scattering rate) at $\omega \sim \Delta_0$  and argued that, depending on parameters, $\text{Im}[\Sigma^R(\omega)]/\omega^2$ either evolves smoothly at $\omega \geq \Delta_0$ or has a peak.  This peak can be detected in ARPES measurements.

The  unconventional behavior that we found  stems from the competition between  different pairing channels. In our case this holds near a nematic QCP, but similar behavior is expected when there is a competition between multiple pairing channels for other reasons, like, e.g., a competition between $s$- and $d$-wave channels in iron-based superconductors 
    \cite{PhysRevLett.109.087001, PhysRevLett.107.117001, HOSONO2015399}. Adapting our computations to a 
 model that only allows for finite number of pairing channels, or, e.g., only selects pairing channels with  even angular momentum yielding singlet superconductivity could be an interesting future project. In addition, a recent Monte Carlo study of a quantum critical metal \cite{grossman2020specific} observed a strong impact of pairing fluctuation on thermodynamic observables (e.g., specific heat), and it would be worthwhile to elucidate how the competition between different pairing  channels shows up in thermodynamics.
 
{As a final remark, we emphasize that our analysis has been restricted to the isotropic case. On a lattice, the nematic form factor $f$ (which we set to one) becomes important, separating hot (where $f \simeq 1$) and cold ($f \ll 1$) regions on the Fermi surface. In the normal state $(\omega \gg \Delta)$, our results carry over to the hot regions. Deep in the superconducting state ($\omega < \Delta$), the gap function will be maximal in the hot regions, and vanish in the cold ones, similar to Ref.\ \cite{PhysRevB.98.220501}. Accordingly, in this limit our results for the scattering rate have to be modified to account for the near-gapless regions. }

\section{Acknowledgments}

We thank  E.~Berg, M.~Hecker, Y.~B.~Kim, A.~Klein, S.-S. Lee and  Y.~Schattner for useful discussions. D.P.\ is grateful to the Max-Planck-Institute for the Physics of Complex Systems Dresden (MPIPKS) for hospitality during the initial stage of this project. The work by A.V.C.\ was supported by the Office of Basic Energy Sciences, U.S. Department of Energy, under award  DE-SC0014402. A.K. was supported by NSF grant DMR-2037654.

\appendix

\section{Particle-hole and particle-particle bubbles in the normal state}
\label{bubblesapp}

The particle-hole bubble of Fig.\ \ref{diagtable}(a) reads
\begin{figure}
\centering
\includegraphics[width=.62\columnwidth]{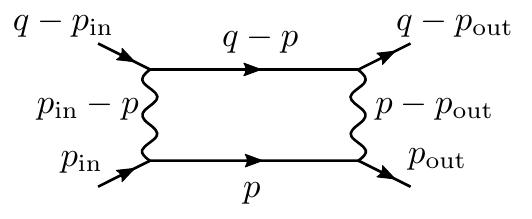}
\caption{Particle-particle bubble $\Pi_{\text{pp}}$ in the form required in Sec.\  \ref{flselfsecrep}. Labels indicate four-momenta.  }
\label{bubblesfig}
\end{figure}
\begin{align}
\Pi_{\text{ph}}(\omega_m,\q) =  N \int_p \frac{1}{i\omega_m^\prime - \xi_\p} \frac{1}{i(\omega_m^\prime + \omega_m) - \xi_{\p + \q}} \ ,
\end{align}
where the factor of $N$ comes from summation over flavors. Evaluation of $\Pi_{\text{ph}}$ is standard (see, e.g., \cite{PhysRevB.74.195126}) and we omit the details here. Performing the $\omega_m^\prime$-integral first, and subsequently the integrals over the linearized dispersion and angular variable, we find
\begin{align}
\Pi_{\text{ph}}(\omega_m,\q)
 = -N\rho + N\rho \cdot \frac{|\omega_m|}{\sqrt{\omega_m^2 + (v_F q)^2}} \ .
\end{align}

Next, we discuss the particle-particle bubble $\Pi_{\text{pp}}$. For the evaluation of Sec.\ \ref{flselfsecrep} it is required in the form shown in Fig.\ \ref{bubblesfig}(b). The internal bubble frequency $\omega_m^\prime$ and momentum modulus $\delta_p = |\p| - k_F$ appears in both interaction lines. However, as discussed in Sec.\ \ref{flselfsecrep}, this dependence only serves to cut off the Cooper logarithm, and can otherwise be neglected. The frequency dependence of the interaction (arising through the Landau damping) leads to the cutoff $\omfl$, while the $\delta_p$ dependence yields a cutoff $v_F M$. Since $\omfl \leq v_F M$ (see discussion above Eq.\ \eqref{sigmaomegaqdef} in the main text), we take into account the frequency cutoff, but integrate over $\delta_p$ in infinite limits.

The interaction lines also depend on the orientation of $\p$, i.e.\ the angular variable $\phi_p$. This dependence needs to be considered in detail. Therefore, as an important auxiliary object we define the  particle-particle bubble with a general angle $\phi_p$ as follows:
\begin{align}
\label{Pippwithphi}
&\Pi_{\text{pp}}(\omega_m, q, \phi_q - \phi_p)   = \\&  \notag k_F\hspace{-.5em}\int\displaylimits_{-\omfl}^{\omfl} \!\frac{d\omega_m^\prime}{2\pi} \int \frac{d\delta_p}{2\pi} \frac{1}{iZ^{-1}(\omega_m - \omega_m^\prime) - \xi_{\q-\p}} \frac{1}{iZ^{-1}\omega_m^\prime - \xi_\p} \ ,
\end{align}
where we also incorporated the one-loop renormalized quasiparticle weight $Z^{-1} = 1+\lambda$.
Linearizing the dispersion, the $\delta_p$ integral is solved by contour integration.
One obtains
\begin{align}
&\Pi_{\text{pp}}(\omega_m, q, \phi_q - \phi_p)   = \\  & \notag  - i \rho  \int_{-\omfl}^\omfl d\omega^\prime \frac{\theta(\omega_m^\prime) - \theta(\omega_m - \omega_m^\prime)}{iZ^{-1}(\omega_m - 2\omega_m^\prime) + qv_F \cos(\phi_q - \phi_p)} \ .
\end{align}
We focus on $\omega_m > 0$; $\omega_m < 0$ yields the same result. We have
\begin{align}
\label{Pippresult}
&\Pi_{\text{pp}}(\omega_m, q, \phi_q - \phi_p)   = \\  & \notag  - i \rho  \int_{\omega_m}^\omfl d\omega_m^\prime \frac{1}{iZ^{-1}(\omega_m - 2\omega_m^\prime) + qv_F \cos(\phi_q - \phi_p)}   \\ \notag &+  i \rho  \int_{-\omfl}^0 d\omega_m^\prime \frac{1}{iZ^{-1}(\omega_m - 2\omega_m^\prime) + qv_F \cos(\phi_q - \phi_p)}\\ &\notag  = Z{\rho} \ln\left( \frac{2Z^{-1} \omfl}{\sqrt{(Z^{-1}\omega_m)^2 + (v_F q \cos(\phi_q - \phi_p))^2}} \right) \ .
\end{align}
The factor $2Z^{-1}$ in the numerator can be neglected with logarithmic accuracy. In addition, since $\phi_q \simeq \pm \pi/2$ plus small corrections of order $\omega_m/v_F q$, we can approximate $\cos^2(\phi_q - \phi_p) \simeq \sin^2(\phi_p)$ (see also App.\ \ref{maincalcdetailssec}), resulting in the static interaction Eq.\ \eqref{Pippmain} of the main text.

For a momentum-independent (BCS-like) interaction, one can integrate over the internal angle and obtains
\begin{align}
\label{PiBCS}
&\Pi_{\text{pp}}^{\text{BCS}} (\omega_m, q) = \\ & \notag Z {\rho} \ln \left(\frac{\omfl}{|Z^{-1}\omega_m| + \sqrt{(v_F q)^2  + |Z^{-1}\omega_m|^2}} \right)  \simeq\\ &  Z{\rho} \ln\left(\frac{\omfl}{v_F q}\right)  - \rho \frac{|\omega_m|}{v_F q} \quad \text{for} \ |\omega_m| \ll v_F q \ .
\notag
\end{align}
\\

\section{One-loop fermion self-energy}
\label{oneloopfermiapp}
Here we detail the evaluation of the one-loop self-energy.
From Fig.\ \ref{diagtable}(b), it is derived as
\begin{widetext}
\begin{align}
\label{Sigma1}
&\Sigma^{(1)}(\omega) = - g\int_q G(\omega_m - \omega_m^\prime, \k - \q) D(\omega_m^\prime, \q)  \simeq \frac{g}{(2\pi)^3}\int d\omega_m^\prime dq  d\phi_q \frac{q}{q^2 + M^2 + g\Pi_{\text{ph}}(\omega_m^\prime,q)} \frac{1}{i(\omega_m - \omega_m^\prime) + v_F q \cos(\phi_q)} \\& = - i \frac{g}{(2\pi)^2} \int d\omega_m^\prime  dq \ \text{sign}(\omega_m - \omega_m^\prime) \frac{q}{\sqrt{(\omega_m - \omega_m^\prime)^2 + (v_F q)^2}}\frac{1}{q^2 + M^2 + g\Pi_{\text{ph}}(\omega_m^\prime,q)} \ . \notag
\end{align}
\end{widetext}
with $\phi_q = \measuredangle(\k,\q)$. To further evaluate Eq.\ \eqref{Sigma1}, we need to assess the typical frequencies and momenta. First, we note that the frequency integral receives contributions from both $|\omega_m^\prime| \lesssim |\omega_m|$ and $|\omega_m^\prime| \gg |\omega_m|$. However, the high-energy region only contributes to the analytical terms in $\Sigma^{(1)}$, as also expected on general grounds. Let us neglect these non-universal terms, and only consider the low-energy contribution $|\omega_m^\prime| \lesssim |\omega|$, i.e., the internal frequencies of interest are set by the external one. We will focus on the limit of small frequencies such that
\begin{align}
\label{omegaFL}
|\omega_m| \ll \omega_{\text{FL}} \equiv \frac{v_F M^3}{N |g| \rho} \ .
\end{align}

This scale can be extracted by equating the Landau damping term $g\Pi_{\text{ph}}(\omega_m,q)$ with $q^2$ for $q = M$. For $\omega_m$ fulfilling Eq.\ \eqref{omegaFL}, $g\piph$ is typically small compared to the bosonic mass $M^2$, and the typical momenta $q$ in Eq.\ \eqref{omegaFL} are given by $q \simeq M$. In the opposite limit $|\omega_m| \gg \omfl$, the Landau damping term dominates over the mass term, and we enter the non-Fermi-liquid regime.

To simplify the evaluation, it is also convenient to require that for typical $q$
\begin{align}
\label{smallomegaapp}
|\omega_m| \ll v_F q \ ,
\end{align}
which is guaranteed if $\omega_m \ll \omfl \leq v_F M$. As a result, we can neglect the frequency dependence under the square-root in Eq.\ \eqref{Sigma1}, and also simplify the Landau damping as $\piph(\omega_m^\prime, q) \simeq N\rho |\omega_m^\prime|/v_F q$. This allows for the approximation
\begin{align}
&\Sigma^{(1)}(\omega_m) \simeq   \\ &\notag  - i \sign(\omega_m) \frac{g}{2\pi^2 v_F } \int_{0}^{|\omega_m|} d\omega_m^\prime dq  \frac{1}{q^2 + M^2 + Ng \rho \frac{\omega_m^\prime}{v_F q}} \ .  \notag
\end{align}
Now $\Sigma^{(1)}(\omega_m) $ can be systematically evaluated by {expanding in the Landau damping}. By simply neglecting the Landau damping, we obtain the term linear in frequency:
\begin{align}
\mathcal{O}(\omega_m)\!: \quad  \Sigma^{(1)}(\omega_m,\k) = - i\lambda \omega_m, \quad \lambda \equiv \frac{g}{4\pi v_F M} \ .
\end{align}

Here, $\lambda$ is an effective dimensionless coupling constant. We do not place any restrictions on it a priori.
The next order term is of order  $\omega_m^2 \ln(\omega_m)$, and requires more careful analysis: the expansion in the Landau damping is strictly legitimate for
\begin{align}
\label{qminApp}
M^2 \gg  Ng \rho \frac{\omega_m^\prime}{v_F q} \Rightarrow q \gg q_\text{min}(\omega_m^\prime) \equiv \frac{Ng \rho \omega_m^\prime}{v_F M^2} ,
\end{align}
while the momenta $q \lesssim q_\text{min}(\omega_m^\prime)$ only contribute to the $\omega_m^2$ term. Since $q_\text{min}(\omega_m^\prime)$ only enters a logarithm, we can replace it by $q_\text{min}(|\omega_m|)$ with logarithmic accuracy (i.e., neglecting $\mathcal{O}(1)$ terms compared to large logarithms). As a result:
\begin{align}
\label{firstlogappendix}
& \mathcal{O}\left(\omega_m^2 \ln(\omega_m)\right)\!:  \quad  \Sigma^{(1)}(\omega_m) =  \\ \notag &  i \sign(\omega_m) \frac{g^2\rho}  {\pi^2 v_F} \int_0^{|\omega_m|} d\omega_m^\prime  \int_{q_\text{min}(|\omega_m|)}^\infty\! dq \frac{1}{(q^2 + M^2)^2} \frac{\omega_m^\prime}{q}  \simeq \\ \notag & i \sign(\omega_m) \frac{g^2 \rho}  {2\pi^2 M^4 v_F} \omega_m^2 \ln\left(\frac{M}{q_\text{min}(|\omega_m|)} \right) = \\ &\notag
 i \text{sign}(\omega_m) \frac{2|\lambda|}{\pi} \frac{\omega_m^2}{\omfl} \ln\left(\frac{\omfl}{|\omega_m|}\right) \ .
\end{align}
Due to the condition \eqref{omegaFL}, the logarithm is indeed large. Evaluation of $\Sigma^{(1)}(\omega_m)$ for $\omega_m \gg \omfl$ can be found e.g.\ in Ref. \cite{PhysRevB.74.195126}.

In Eq.\ \eqref{firstlogappendix}, internal momenta $q$ are small in a logarithmic sense:
one can think about $q$ as $q \propto \omega_m^\alpha$ with  $0< \alpha \ll 1$, such that $v_F q \gg \omega_m$, but at the same time $v_F q \ll \omfl, v_F M$ is much smaller than the UV scales in the problem. Thus, given an external wave-vector $\k$, the internal fermions are fixed to $\pm \k$, and momentum variations around these points are small in the external $\omega_m$; the fermions almost move along a line.

\section{Three-loop calculations}
\label{threeloopApp}

To evaluate the three-loop diagrams, it is convenient to use patch coordinates \cite{Sachdev2011},  expanding the fermionic dispersion around the points $\k,-\k$ as
\begin{align}
\label{patchdispersion}
\xi(\pm \k + \q) = \pm v_Fq_x + q_y^2/(2m) \ .
\end{align}
First, we compute the three-loop diagram in the forward scattering channel of Fig.~\ref{planar3loop}(a), with $k = (\omega_m, \k_F, 0),\  q = (\omega_m^\prime,
q_x, q_y), q_i = (\omega_{im}, q_{ix}, q_{iy}), i = 1,2$. It will be denoted $\Sigma^{(3)}_\text{a}(\omega_m)$. This diagram reads
\begin{align}
\label{Sigma3fwapp}
&\Sigma^{(3)}_\text{a}(\omega) = g^3 N \int_{q_1,q_2,q} G(k- q) G(k + q_1 - q) \times  \\ & G(k + q_1)  G(k + q_2 - q) G(k + q_2) D(q_1) D(q_2) D(q_2 - q_1) \ .   \notag
\end{align}
We will use a simplified boson propagator:
\begin{align}
D(q_1) \simeq - \frac{1}{ M^2 + q_{1y}^2 + \frac{g \rho N}{v_F} \frac{|\omega_{1m}|}{|q_{1y}|}} \  ,
\end{align}
which holds for $M \ll k_F$ where typically $q_{1x} \ll q_{1y}$. The $q_{1x}$ integral in \eqref{Sigma3fwapp} can then be computed by contour integration, and leads to \begin{widetext}
\begin{align} \notag
I_{1x} &\equiv \int \frac{dq_{1x}}{2\pi} \frac{1}{iZ^{-1}(\omega_m + \omega_{1m}) - v_F q_{1x} - q_{1y}^2/{2m} }  \times \frac{1}{iZ^{-1}(\omega_m + \omega_{1m} - \omega_m^\prime) - v_F(q_{1x} - q_x) - (q_{1y} - q_y)^2/{2m}} \\ &= - \frac{i}{v_F} \frac{\theta(\omega_m + \omega_{1m}) - \theta(\omega_m + \omega_{1m} - \omega_m^\prime)}{-Z^{-1}i\omega_m^\prime + v_F q_x + q_{1y}^2/2m - (q_{1y} - q_y)^2/2m}\ . \label{firststepfs}
\end{align}
The $q_{2x}$ integral is performed alike. Now performing the $q_x$ integral
\begin{align} \notag
I_x &\equiv   \int \frac{dq_x}{2\pi}   \frac{1}{-iZ^{-1}\omega_m^\prime + v_F q_x + q_{1y}^2/2m - (q_{1y} - q_y)^2/2m} \\ & \notag \times\frac{1}{-iZ^{-1}\omega_m^\prime + v_F q_x + q_{2y}^2/2m - (q_{2y} - q_y)^2/2m}\times\frac{1}{iZ^{-1}(\omega_m - \omega_m^\prime) + v_F q_x - q_y^2/2m} \\  & = -\frac{i}{v_F}\frac{\theta(\omega_m - \omega_m^\prime) - \theta(-\omega_m^\prime)}{-iZ^{-1}\omega_m + q_y^2/2m + q_{1y}^2/2m - (q_{1y} - q_y)^2/2m }  \times  \frac{1}{-iZ^{-1}\omega_m + q_y^2/2m + q_{2y}^2/2m - (q_{2y} - q_y)^2/2m }  \ . \label{omegaqstep}
\end{align}

Notice the partial cancellation of the curvature terms; only mixed terms $q_{iy} q_y$ remain. Let $\omega_m > 0$. The step-function in Eq. \eqref{omegaqstep} restricts $\omega_m^\prime \in (0,\omega_m)$, and the two step-functions in \eqref{firststepfs} then imply $\omega_{im} \in (-\omega_m, \omega_m^\prime - \omega_m)$. Combining the results obtained so far:
\begin{align}
&\Sigma^{(3)}_\text{a}(\omega_m) = \\ & \notag i\frac{g^3N}{(2\pi)^6v_F^3} \int_0^\omega d\omega_m^\prime \int\displaylimits_{-\omega_m}^{\omega_m^\prime - \omega_m} d\omega_{1m} d\omega_{2m} \int dq_y dq_{1y} dq_{2y} \frac{1}{-iZ^{-1}\omega_m + q_{1y} q_y/m} \frac{1}{-iZ^{-1}\omega_m + q_{2y} q_y/m} D(q_1) D(q_2) D(q_2 - q_1) \ .
\end{align}
The $q_y$-integral comes from poles $q_y = -iZ^{-1}\omega_m m / q_{iy}$, with $q_{iy} = \mathcal{O}(M)$. I.e., it does not come from $|\q_i|$ which are small in $\omega$, as stated in the main text. It yields
\begin{align} \notag
\Sigma^{(3)}_\text{a}(\omega_m) &=  i\frac{g^3N}{(2\pi)^5v_F^3} \int_0^{\omega_m} d\omega_m^\prime \int_{-\omega_m}^{\omega_m^\prime - \omega_m} d\omega_{1m} d\omega_{2m} \int  dq_{1y} dq_{2y} \frac{m(\theta(q_{1y}) - \theta(q_{2y}))}{Z^{-1}\omega_m(q_{2y} - q_{1y})} D(q_1) D(q_2)  D(q_2 - q_1) \\&= -  i \frac{g^3Z N m}{(2\pi)^5v_F^3}  \frac{2}{\omega}  \int_0^{\omega_m} d\omega_m^\prime \int_{-\omega_m}^{\omega_m^\prime - \omega_m} d\omega_{1m} d\omega_{2m} \int_0^\infty dq_{1y} dq_{2y} \frac{1}{q_{1y} + q_{2y} } D(q_1) D(q_2) D(\omega_{1m} - \omega_{2m}, q_{1y}+q_{2y}) \ .
\end{align}
\end{widetext}
To leading order in $\omega_m$ we can neglect the dynamics (Landau damping) in the Boson propagator. Computing the trivial frequency integral, and rescaling the remaining momenta $y_i \equiv q_{iy}/M$, we arrive at
\begin{align}
&\Sigma^{(3)}_\text{a}(\omega_m)  = {i \text{sign}(\lambda)  \frac{1}{\pi^2} \frac{\lambda^2}{(1+ \lambda)} \frac{\omega_m^2}{\omfl}\times\text{const.}}\ , \\
&\text{const.} = \frac{2}{3}\!\int_0^\infty\!dy_1 dy_2  \frac{1}{y_1 + y_2} \frac{1}{1 + y_1^2} \frac{1}{1 + y_2^2}\frac{1}{1 + (y_1 + y_2)^2} \notag \\&= 0.56329\hdots \notag .
\end{align}
Since there is no linear contribution, we write $\Sigma^{(3)}_\text{a}(\omega_m) = \Sigma^{(3)}_{\text{a},\omega^2}(\omega_m)$ as in Eq.\ \eqref{forwardresult3loop} of the main text.

The $\omega_m^2$ part of the remaining three-loop diagrams is more conveniently computed by first evaluating the dressed vertices (blue boxes in Fig.\ \ref{planar3loopvertices}). For the planar diagram in the backscattering channel (Fig.\ 5(b)), we only need the  vertex as function of $q_y$ and can set the other variables to zero (see also Sec.\ \ref{flselfsecrep}). Thus, we have
\begin{align} \notag
&\Gamma_{\text{bs}}^{(1)} (q_y) =  - g^2 \int_{q_1} D(q_1)^2 \ \frac{1}{iZ^{-1}\omega_{1m} - v_F q_{1x} - q_{1y}^2/2m}  \\ &  \times \frac{1}{-iZ^{-1}\omega_{1m} - v_F q_{1x} - (q_y - q_{1y})^2/2m} \ .
\end{align}
The $q_{1x}$-integral yields
\begin{align}
\label{Gammabs1nocurve}
&\Gamma_{\text{bs}}^{(1)} (q_y) =  g^2\frac{i}{v_F} \int \frac{dq_{1y} d\omega_{1m}}{(2\pi)^2} \text{sign}(\omega_{1m}) \times \\ & \notag \frac{1}{-2iZ^{-1}\omega_{1m} + {(q_{1y})^2}/{2m} - (q_y - q_{1y})^2/2m}  \times \\ & \notag \frac{1}{\left((q_{1y})^2 + M^2 + \frac{N\rho g}{v_F} \frac{|\omega_{1m}|}{|q_{1y}|} \right)^2} \ .
\end{align}
The $\omega_{1m}$ integral is logarithmic, and the boson propagators just set the cutoff; since the typical value of $q_{1y}$ is $M$, the cutoff becomes $\omfl$. Then we find, with logarithmic accuracy:
\begin{align}
&\Gamma_{\text{bs}}^{(1)} (q_y) \simeq -\frac{ Z g^2}{v_F (2\pi)^2} \int dq_{1y} \frac{1}{(q_{1y}^2 + M^2)^2}  \\ & \times\ln\left(\frac{\omfl}{|q_{1y} q_y/m - q_y^2/2m|} \right) \notag \ .
\end{align}
In the logarithm, we can replace the argument by a typical value. We will need $q_y \ll M$, therefore we can approximate it as $q_{1y} q_y /m \simeq M q_y /m = q_y v_F \epsilon$, with $\epsilon = M/k_F$. Then the integral results in
\begin{align}
&\Gamma_{\text{bs}}^{(1)} (q_y) \simeq - \frac{g}{M^2} \times \frac{1}{2} \frac{\lambda}{1+\lambda} \ln\left(\frac{\omfl}{v_F q_y \epsilon} \right) \ .
\end{align}
Inserting this vertex into Eq.\ \eqref{sigmaomegaqdef_new} (with variable $z = v_F q \epsilon$, see also discussion below Eq.\ \eqref{Sigmap2}) immediately yields $\Sigma_{\text{b},\omega^2}^{(3)}$ from Eq.\ \eqref{backscatteringhere}.

To evaluate the non-planar diagram of Fig.\ \ref{planar3loopvertices}(c), we likewise evaluate the vertex, to be denoted $\Gamma_{\text{np}}$. Here we can set external variables to zero:
\begin{align}
\label{Gammanpfirst}
&\Gamma_{\text{np}} \equiv - g^2 \int_{q_1} D(q_1)^2 \times \\ \notag &   \frac{1}{iZ^{-1}\omega_{1m} + v_F q_{1x} - q_{1y}^2/2m} \frac{1}{iZ^{-1}\omega_{1m} - v_F q_{1x} - q_{1y}^2/2m}     \\ \notag & =   \frac{ig^2}{(2\pi)^2 v_F} \int d\omega_{1m} dq_{1y}  \frac{\text{sign}(\omega_{1m})}{2iZ^{-1}\omega_{1m} - q_y^2/m} D(q_1)^2 =  \\ \notag &  \frac{g^2}{(2\pi)^2 v_F} \int d\omega_{1m} dq_{1y}  \frac{2Z^{-1}|\omega_{1m}|}{4(Z^{-1}\omega_{1m})^2 + \frac{q_{1y}^4}{m^2}} \\ & \notag \times \frac{1}{\left(M^2 + q_y^2 + \frac{g \rho N}{v_
F}\frac{|\omega_{1m}|}{|q_{1y}|}\right)^2} \ .   \notag
\end{align}
To extract the limit $N \rightarrow \infty$, we introduce dimensionless variables as
\begin{align}
y = \frac{q_{1y}}{M \alpha^2}, \quad x = \frac{Z^{-1}\omega_{1m}m}{ M^2\alpha}, \quad \alpha = \frac{m}{\rho N \lambda}  \ .
\end{align}
This yields
\begin{align}
\Gamma_{\text{np}} =  \frac{1}{2\pi} \frac{g}{M^2} \lamt \alpha \times  2 \int_0^\infty dx dy \frac{x}{4x^2 + y^4} \frac{1}{1+y^2\alpha^2 + \frac{x}{y}} \ .
\end{align}
In the limit $N \rightarrow \infty$ where $\alpha \rightarrow 0$, the last factor has a well-defined limit and one obtains
\begin{align}
\Gamma_{\text{np}} \rightarrow    \frac{1}{2\pi} \frac{g}{M^2} \lamt \alpha =  \frac{1}{2\pi} \frac{g}{M^2} \frac{1}{1+\lambda}\frac{m}{N\rho} = \frac{g}{M^2} \frac{1}{(1+\lambda)N} \ .
\end{align}
Inserting this instead of the bare vertex at zero energy-momentum $(-g/M^2)$ into the one-loop diagram, one obtains Eq.\ \eqref{backresult3loop_1}.

\section{Stucture of planar diagrams}
\label{planarstructureapp}

\begin{figure}
\centering
\includegraphics[width=\columnwidth]{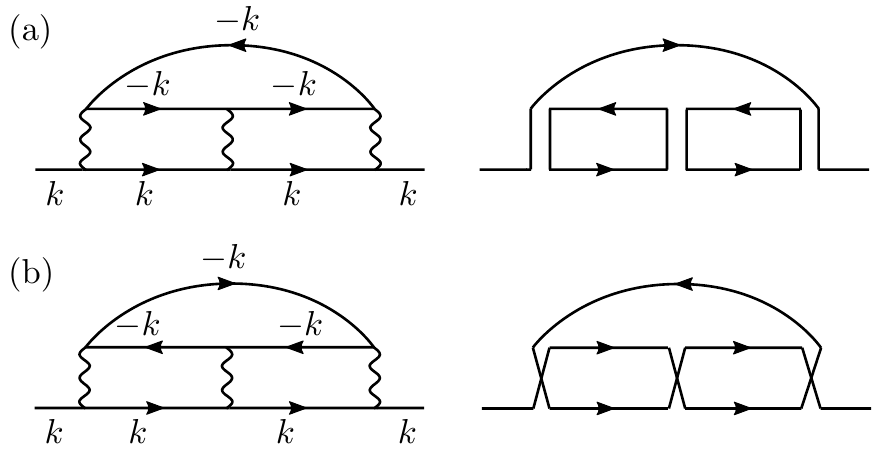}
\caption{{Double line representation of a planar (a) and non-planar diagram (b). The planar diagram is free of crossings. Fig.\ adapted from Ref.\ \cite{metlitski2010quantum}.} }
\label{planarGraphical}
\end{figure}

Let us first recall the large-$N$ analysis of Refs.~\cite{PhysRevB.73.045128, metlitski2010quantum,PhysRevB.82.045121} at the QCP. These authors rescale the bare coupling constant $g \rightarrow g/N$. Then the Landau damping $\Pi$ is  $\mathcal{O}(1)$, while the one-loop quantum critical self-energy $\Sigma^{(1)}(\omega) = (\omega_{\text{IN}})^{1/3} \omega^{2/3}$ of the fermions picks up a factor $1/N$. Naively, all diagrams in Fig.\ \ref{planar3loop} then have the power $1/N^2$ and are subleading compared to $\Sigma^{(1)}(\omega)$. However, the two planar diagrams (Fig.\ \ref{planar3loop} (a),(b)) contain a ``singular manifold'' of dimension two where all internal fermions are on the Fermi surface -- if the momentum $\q$ in the diagram is set to zero, there are two free parameters for $\k + \q_1,\k + \q_2$ to be on the Fermi surface \cite{PhysRevB.80.165102}. By counting the strength of poles of fermionic propagators, one sees that this zero-energy manifold would induce a IR singularity were it not for the frequency dependence of the propagators. Since the leading frequency-dependence comes from $\Sigma$, which scales a $1/N$ as stated above, the diagrams acquire at least one additional power of $N$ \footnote{It turns out that the backscattering diagram is even of order $\mathcal{O}(N^0)$ due to a UV singularity arising in the quantum-critical limit \cite{metlitski2010quantum}}. A graphical way to recognize planarity is by replacing the wavy boson lines by two fermion lines, and inverting the fermion line direction at the $(-\k)$ patch \cite{PhysRevB.80.165102, metlitski2010quantum}. {As visualized in Fig.\ \ref{planarGraphical}, in this double-line representation only the planar diagrams  be drawn on a sphere without any crossings; the non-planar diagrams can be ``untwisted'' on higher-genus surfaces only}. The additional powers of $N$ which the diagram contains beyond the naive power counting are determined by the number of ``single-line'' loops in this representation.

From another perspective, the planar diagrams are distinguished by  a  \textit{cancellation of Fermi surface curvature} \cite{metlitski2010quantum}. In the patch theory dispersion $v_F k_x + k_y^2/2m$ we regard $v_F$ as fixed, and the Fermi surface curvature is given by $1/m$. The cancellation of the curvature occurs in the non-Fermi-liquid and Fermi-liquid cases alike, and is explicit in the three-loop evaluations of App.\ \ref{threeloopApp}: in the forward-scattering  planar diagram, after the integrals over the $x$-momenta are taken, the terms $q_y^2, q_{1y}^2, q_{2y}^2$ cancel out in Eq.\ \eqref{omegaqstep}; only mixed terms remain. Such cancellation also occurs in the planar backscattering vertex $\Gamma_{\text{bs}}^{(1)}$, Eq.\ \eqref{Gammabs1nocurve}. By contrast, in the planar backscattering vertex $\Gamma_{\text{np}}$ there is no cancellation. see Eq.\ \eqref{Gammanpfirst}. As a result, this vertex incures an additional factor $(m/N\rho)$ compared to the planar one.

For a general diagram, one can expect that the (partial) cancellation of curvature is an equivalent criterion to the ``singular manifold'' introduced in Ref.\ \cite{PhysRevB.80.165102}: as shown in this reference for the one-patch model, by assigning momenta such that all fermions are on the Fermi surface, the momenta of propagators connected to the external legs in the double-line representation are fixed. However, all fermions in the closed ``single-line''-loops carry the same unfixed momentum $\k + \p$. By integrating over the $p_x$-component, the curvature in the single-line loop cancels as long as one can neglect the $p_x$-dependence of the bosons. A similar argument applies if the fermions are part of opposite patch and have momentum $-\k - \p$, which results in the same pole structure of the propagator.
In conclusion, it is thus expected that the planar diagrams are leading in the $\mathcal{O}(1/N)$ expansion not only at the QCP, but also in a Fermi liquid.  \\

\section{Detailed evaluation of $\text{Im}[\Sigma_{\text{bs}}]$}
\label{maincalcdetailssec}

Let us fill in the details for the evaluation of $\text{Im}[\Sigma_{\text{bs}}]$. Consider first an interaction line in Fig.\ \ref{maindiagram}, say $D(p_1 - p_2)$: 
\begin{widetext}
\begin{align}
\label{interactionformapp}
&D(p_1 - p_2) =   -g \left[M^2 + (\p_1 - \p_2)^2 + N g \rho \frac{|\omega_{1m} - \omega_{2m}|}{ v_F |\p_1 - \p_2| }\right]^{-1}  =   -g \bigg[ M^2 + 4k_F^2 \sin^2\!\left(\frac{\phi_{1} - \phi_{2}}{2}\right)  \\ & \notag + 4k_F(\delta_{p_1} + \delta_{p_2}) \sin^2\!\left(\frac{\phi_{1} - \phi_{2}}{2}\right) + \delta_{p_1}^2 + \delta_{p_2}^2 - 2\delta_{p_1} \delta_{p_2} \cos(\phi_{2} - \phi_{1})
+ N g \rho \frac{|\omega_{1m} - \omega_{2m}|}{ v_F |\p_1 - \p_2| }\bigg]^{-1} ,
\end{align}
\end{widetext}
 where $\delta_{p_i} = |\p_i| - k_F$, and angles are measured relative to $\k$. Recall the parameter $\epsilon = M/k_F \ll 1$. Eq.\ \eqref{interactionformapp} and the remaining interaction lines in the diagram restrict the typical angles to be $\phi_i \lesssim \epsilon$. Thus, in each Cooper bubble the typical momentum along the Fermi surface at the point $\k$, given by $|\p_i| \sin(\phi_i)$, is of order $M$.
 
Fortunately, the detailed dependence on $\omega_{im}, \delta_{p_i}$ in Eq.\ \eqref{interactionformapp} does not have to be taken into account, since it is slow compared to the singular logarithmic dependence in the Cooper bubbles. However, the interaction lines set the UV cutoff on the logarithmic Cooper integral: $|\delta_{i}| < M$ and $|\omega_{im}| < \omfl$.
For small $M$, the latter cutoff is more important, and it is thus sufficient to evaluate the particle-particle bubble $\Pi_{\text{pp}}$ restricting the internal frequency integral only. This is done in App.\ \ref{bubblesapp}, Eq.\ \eqref{Pippresult},  with the result
\begin{align}
\label{Pippmaintext}
&\Pi_{\text{pp}}(\omega_m^\prime, q, \phi_q - \phi_i) = \\& \notag Z\rho \ln\left( \frac{\omfl}{\sqrt{(Z^{-1}\omega_m^\prime)^2 + (v_F q \cos(\phi_q - \phi_i))^2}} \right) ,
\end{align}

where we did not perform the angular integral yet since the angular dependence of the interaction lines is crucial per the above . Eq.\ \eqref{Pippmaintext} can be simplified by inserting a typical value of $\phi_q$: the integral over $\phi_q$ is of the form
 \begin{align} \int d\phi_q \frac{1}{iZ^{-1}(\omega_m - \omega_m^\prime) + v_F q \cos(\phi_q)} \hdots
\end{align}
In the most important limit $\omega_m, \omega_m^\prime \ll v_F q$, it is dominated by angles $\phi_q = \pm \pi/2$ up to corrections of order $\omega_m/v_Fq$, and we can therefore approximate
\begin{align}
\label{Pippmaintextsin}
&\Pi_{\text{pp}}(\omega_m^\prime, q, \phi_q - \phi_i) \simeq \\& \notag Z\rho \ln\left( \frac{\omfl}{\sqrt{(Z^{-1}\omega_m^\prime)^2 + (v_F q \sin(\phi_i))^2}} \right).
\end{align}
Let us recall the form of $\Pi_\text{pp}$ for a momentum-independent interaction (calling it BCS-like). Then the angular integral can simply be taken, and one finds \cite{PhysRevB.68.155113, *PhysRevB.74.079907}
\begin{align} \notag
&\Pi_{\text{pp}}^{\text{BCS}} (\omega_m^\prime, q) =  \\ & \notag  Z\rho \ln \left(\frac{\omfl}{|Z^{-1}\omega_m^\prime| + \sqrt{(v_F q)^2  + |Z^{-1}\omega_m^\prime|^2}} \right) \simeq   \\ &  Z \rho \ln\left(\frac{\omfl}{v_F q}\right)  - \rho \frac{|\omega_m^\prime|}{v_F q} \ . \label{PiBCS}
\end{align}
The dynamical term is nothing but the Landau damping up to a  factor of $N$ (since our $\Pi_{\text{pp}}$ does not involve a flavor sum).
As seen from Eq.\ \eqref{Pippmaintextsin}, this term arises from typical values  $\phi_i \simeq \omega_m^\prime/v_Fq$ or $\simeq \omega_m^\prime/v_F q+\pi$. With our momentum-dependent interaction, $\phi_i \simeq \pi$ can be neglected as involves a large momentum transfer. On the other hand, the static Cooper logarithm in \eqref{PiBCS} comes from $\mathcal{O}(1)$ values of the angle. Put together: the Landau damping in the Cooper bubble comes from internal momenta  along the Fermi surface which are small in $\omega_m$, while the Cooper logarithm comes from much larger momenta of order $M$.

These considerations outline the following strategy for evaluation of $\Sigma_{\omega^2, \text{bs}}$:
We need to select the Landau-damping part from one particle-particle bubble, with a factor $1/2$ compared to Eq.\ \eqref{PiBCS}; since the angle $\phi_i$ corresponding to this bubble is small in $\omega_m/v_Fq$, we can set it to zero in the interactions \eqref{interactionform}. From the remaining bubbles, we only take the static Cooper logarithm, setting $\omega_m^\prime = 0$ there. The angle dependence in the Cooper logarithm can be replaced by a typical value, $\sin(\phi_i) \simeq  \epsilon$, and\ the Cooper logarithms become
\begin{align}
L(z) \equiv  \ln\left(\frac{\omfl}{z} \right) \ ,
\label{Cooperlog}
\end{align}
where
$ z = v_F q \epsilon$. All Cooper diagrams can then be summed by
 selecting a cross-section in which we take the Landau damping piece from the particle-particle bubble, and
summing Cooper logarithms to the left and right of it, corresponding to the evaluation of the dressed backscattering amplitude $\Gamma_{\text{bs}}(z)$. Note that is crucial to retain the Landau-damping piece once, without this we would miss the universal contribution with
 $\omega_m^2$ in front of the logarithmic dependence. 
 
 Proceeding as in the main text, one arrives at Eq.\ \eqref{Sigmap2}: 
 \begin{align} 
 \label{firstzapp}
 &\Sigma_{\omega^2, \text{bs}}(\omega_m) = i \sign(\omega_m) \frac{\lambda}{\pi} \frac{\omega^2_m}{\omfl}   \int_{\omega_m}^{\omfl}  \frac{dz}{z}
 \left[\frac{M^2}{g} \Gamma_{\text{bs}} (z)\right]^2 
 \end{align} 
 
In this equation, the integral boundaries have been chosen as
follows: The upper boundary is fixed in such a way that the form of the Cooper logarithms remains valid; as per Eq.\ \eqref{Pippmaintextsin}, this requires $z  \ll \omfl$. Likewise, we can only neglect $\omega_m^\prime \sim \omega_m$ in the Cooper logarithms if $\omega_m \ll z$.  This sets the lower boundary of integration over $z$.

From Eq.\ \eqref{firstzapp}, the final results for both repulsive and attractive interactions can be obtained by substituting $x = |\lamt| \ln(\omfl/z)$, which results in 

\begin{align}
\label{fxintegral}
\Sigma_{\omega^2,\text{bs}} = &i \sign(\omega_m) \frac{\omega^2_m}{\pi \omfl} \frac{|\lambda|}{|\lamt|}\int_0^\ell dx \left(\frac{\ln( 1 \pm x)}{x} \right)^2
\end{align}

where the $+ (-)$ sign corresponds to repulsive (attractive) interactions, and $\ell = |\lamt| \ln(\omfl/|\omega_m|)$. Evaluating the integrals we obtain Eqs.\ \eqref{Sigmawithxrep}, \eqref{sigmaattleading} of the main text. We note that the integral for attractive interactions and integral boundary $\ell/(1+\ell)$ can be mapped on the integral for repulsive interactions by substituting $y = x/(1-x)$, which shows  relation \eqref{fssym} of the main text.

\section{$\Gamma_{\text{bs}}$ beyond leading order}
\label{Tmatrixapp}

\begin{figure}
\centering
\includegraphics[width=1.05\columnwidth]{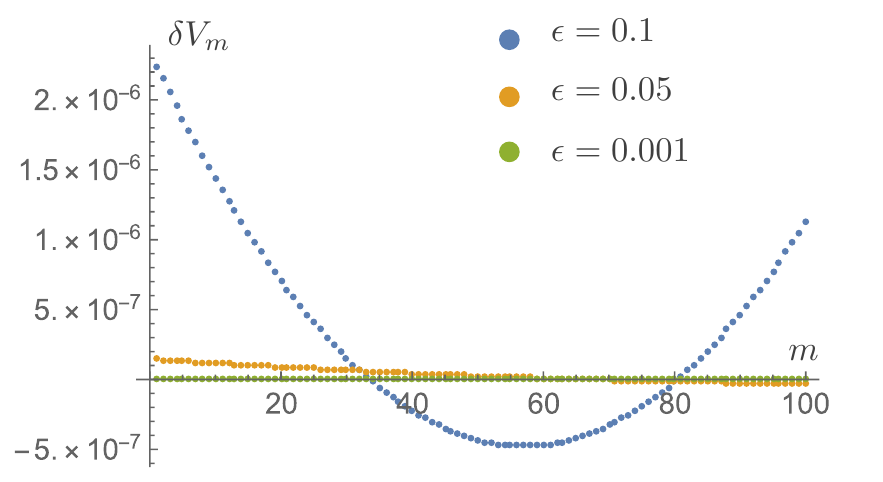}
\caption{Relative error $\delta V_m$ of $V_m^{(2)}$, Eq.\ \eqref{Vl2}, as function of $m$, for three different values of $\epsilon$. $\delta V_m$ is defined as $\delta V_m = V_m/V_m^{(2)} -1$, with $V_m$ obtained from numerical integration.  }
\label{Vlcheck}
\end{figure}

The backscattering amplitude $\Gamma_{\text{bs}}$ has been defined in Eq.\ \eqref{Tfirst} as a function of the $z$-dependent logarithm $L(z)$, with $z = v_F q \epsilon$. Suppressing the $z$-dependence, we can rewrite this definition as
\begin{widetext}
\begin{align}
&\Gamma_{\text{bs}} = - \frac{1}{\rho} \sum_{n=0}^\infty  \left(L\right)^n \int_0^{2\pi} \frac{d\phi_1}{2\pi} \frac{d\phi_2}{2\pi} \hdots \frac{d\phi_n}{2\pi} V(\phi_1) V(\phi_2 - \phi_1) \hdots V(\phi_n - \phi_{n-1}) V(\phi_n) \ , \quad V(\phi) \equiv \frac{2\epsilon \lamt}{\epsilon^2 + 4\sin^2(\phi/2)} \ .
\end{align}
\end{widetext}
Introducing the angular momentum decomposition, $\Gamma_\text{bs} $ can be reformulated as in Eq.\ \eqref{remsum}:

\begin{align} \label{Tdecapp}
 &\Gamma_\text{bs}  = \\ \notag  &- \frac{1}{\rho} \sum_m V_m \sum_{n = 0}^\infty \left[ V_m L\right]^{n}  = - \frac{1}{\rho} \frac{V_0}{1-V_0 L} - \frac{2}{\rho} \sum_{m = 1}^\infty \frac{V_m}{1- V_m L},
\end{align}
with $V_m$ the angular momentum components of $V$, which are defined as
\begin{align}
\label{Vldefapp}
V_m = \int_0^{2\pi} \frac{d\phi}{2\pi} V(\phi) \exp(-i\phi m) = V_{-m} \ .
\end{align}
To compute $\Gamma_\text{bs}$, $V_m$ is needed as an input. Let $m > 0$. For fixed small $m$, $V_m$ can be evaluated explicitly:
\begin{align}
\label{lowVl}
&V_0 = \lamt \frac{2}{\sqrt{4 + \epsilon^2}}  \\ & \notag
V_1 = \lamt\left(\frac{\epsilon ^2+2}{\sqrt{\epsilon ^2+4}}-\epsilon\right)  \\ & \notag
V_2 = \lamt \left(\frac{2}{\sqrt{\epsilon ^2+4}}-\epsilon  \left(\epsilon ^2-\sqrt{\epsilon ^2+4} \epsilon
   +2\right)\right) \hdots
\end{align}
To obtain a useful approximation to $V_m$ for general $m$ and small $\epsilon$, we can evaluate \eqref{Vldefapp} by extending boundaries to infinity and replacing $4 \sin^2(\phi/2) \simeq \phi^2$. This yields
\begin{align}
\label{Vl1app}
V_m^{(1)}  \simeq \tilde \lambda \exp(-\epsilon m) \ .
\end{align}
Comparison to Eq.\ \eqref{lowVl} indicates that this expression is valid up to order $\mathcal{O}(\epsilon^2)$.
From Eq.\ \eqref{lowVl} one can also make an educated guess for the next order terms in $V_m$:
\begin{align}
\label{Vl2}
V_m^{(2)} = \tilde \lambda \exp(-\epsilon m ) \left( 1 - \epsilon^2/8 + \frac{m}{24} \epsilon^3 + \hdots \right) .
\end{align}
This improved approximation is numerically checked in Fig.\ \ref{Vlcheck}, and has a relative error of $\lesssim 10^{-6}$ even for $\epsilon = 0.1$. As a takeaway point, we conjecture that already  $V_m^{(1)}$ from Eq.\ \eqref{Vl1app} reproduces the leading (i.e., largest) power in $m$ at each order in $\epsilon$, which will prove enough for our purposes.

We return to Eq.\ \eqref{Tdecapp} and consider the sum over higher angular momenta:
\begin{align}
\frac{2}{\rho} \sum_{m = 1}^\infty \frac{V_m}{1- V_m L} = \frac{2}{\rho} \sum_{m = 1}^\infty \frac{1}{\lamt/V_m - 1 + (1 - \lamt L)} \ .
\end{align}
To evaluate the sum, we can apply the Euler-Maclaurin formula in the form \cite{abramowitz1972handbook}
\begin{align}
\label{EMLform}
\sum_{m = 1}^\infty f(m) = \int_1^\infty dm f(m) + \frac{1}{2} f(1) + R_f ,
\end{align}
where $R_f$ is a rest term; we will numerically check that $R_f$ can be neglected below. We approximate $V_m =\tilde \lambda \exp(-m \epsilon)$ and therefore use
\begin{align}
f(m) = \frac{1}{\exp(m \epsilon) -1 + (1 - \lamt L)} \ .
\end{align}
Then $f(1) \simeq 1/(\epsilon + (1 - \lamt L))$, which will be subleading. The integral in Eq.\ \eqref{EMLform}  be computed directly, but it is more transparent to get the result by expanding $f$ in $\epsilon$, which is approximately valid as long as $m \lesssim 1/\epsilon$:
\begin{align}
\label{logeps}
&\int_1^{1/\epsilon} dm \frac{1}{\epsilon m + \frac{1}{2} (\epsilon m)^2 + (1 - \lamt L)} \overset{x = m\epsilon}{=}  \\ & \notag \frac{1}{\epsilon} \int_{\epsilon}^{1} dx \frac{1}{x} \frac{1}{1 + x/2 + (1 - \lamt L)/x} \simeq \\ \notag & \frac{1}{\epsilon} \ln\left(\frac{1}{(1- \lamt L) + \epsilon}\right) , \notag
\end{align}
where the upper cutoff of the logarithm is determined up to factors of order $\mathcal{O}(1)$. The last form of the integral shows that knowledge of the leading order coefficients in $m$, $\epsilon$ for $V_m$ is indeed sufficient: for instance, in Eq.\ \eqref{Vl2} it was claimed that the $\epsilon^2$-term in $V_m$ actually reads $\epsilon^2(m^2/2 - 1/8)$. However, the extra term $1/8 \epsilon^2$ would only lead to an $\mathcal{O}(\epsilon)$ correction to the upper cutoff in Eq.\ \eqref{logeps}, which can be neglected.

\begin{figure}
\centering
\includegraphics[width=\columnwidth]{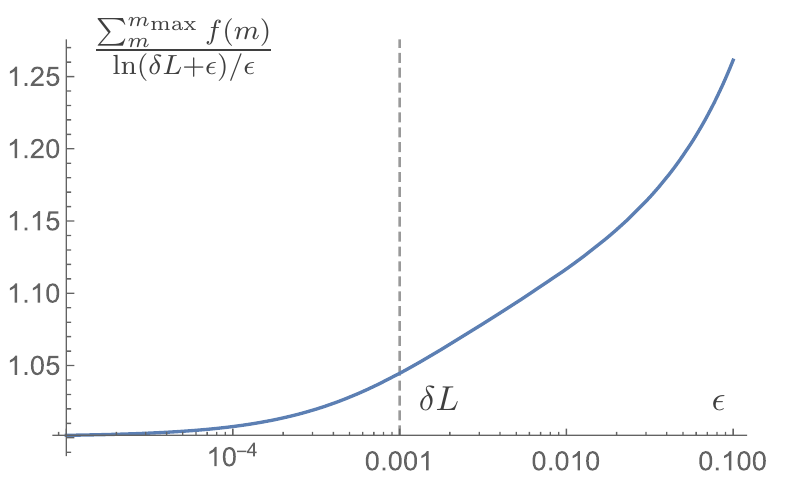}
\caption{Numerical check of the Euler-MacLaurin approximation for $\delta L = 1- \lamt L = 0.001$. The sum is truncated at $m_{\text{max}} = 10^3  \left \lfloor{1/(\epsilon + \delta L)} \right \rfloor$. }
\label{EMLfig}
\end{figure}

In Fig.\ \ref{EMLfig}  we compare the approximate result \eqref{logeps} to direct numerical evaluation of the appropriately truncated sum in \eqref{EMLform}. We find very good agreement as $\epsilon \rightarrow 0$, indicating that the remainder term $R_f$ can be disregarded.

We can combine  Eq.\ \eqref{logeps} with the $s$-wave part and set $V_0 \simeq \lamt$ (which holds up to $\mathcal{O}(\epsilon^2)$). Dividing the logarithm by $\lamt L(z)$ to match the high-energy form of $\Gamma_{\text{bs}}(z)$,
 we obtain Eq.\ \eqref{Tnextto} of the main text.

\section{Self-energy in the superconducting state: contribution of higher angular momentum channels}
\label{Imsigma1app}

To obtain Eq.~\eqref{Imsigma3delta} in the main text, we need to re-evaluate the one-loop diagram of Fig.\ \ref{diagtable}(b) in the superconducting state. This is easiest done in the spectral representation, where, for $\omega > 0$:
\begin{align}
\label{spectralrep}
&\text{Im}\left[\Sigma_\text{bs}^R(\omega)\right] =(-  g) \times  \\ & \notag  \int \frac{d\q}{(2\pi)^2} \int_0^\omega \frac{d\omega^\prime}{\pi} \text{Im}\left[G^R(\omega - \omega^\prime, \k +\q)\right] \text{Im} \left[ D^R(\omega^\prime, \q) \right] \ .
\end{align}
To evaluate this formula, we need to find $\text{Im} \left[ D^R(\omega^\prime, \q) \right]$ (the RPA dressed interaction), which requires knowledge of the particle-hole bubble $\Pi_{\text{ph}}$. In the superconducting state, it becomes $GG - FF$ (see, e.g., Sec.\ 10.2.4 of Ref.\ \cite{mahan2013many}), with propagators as in Eq.\ \eqref{Gmeanfieldmain}. In the Matsubara domain, after evaluating lengthy but straightforward momentum integrals, one arrives at
\begin{align}
& \notag \Pi_{\text{ph}}(\omega_m^{\prime}, \q) = N \int_p  \left[ G(p + q) G(p) - F(p+q) F(p) \right]=
\\ & \notag  N \rho \int_0^\infty d\tilde\omega_m \frac{\Omega_{+} \Omega_{-} + (\omega_m^\prime)^2/4 - (\tilde\omega_m)^2 - \Delta_0^2}{ \Omega_{+} \Omega_{-} \left(\left(\Omega_{+} + \Omega_{-}\right)^2 + (v_F q)^2 \right)^{1/2}} \ , \\ &\Omega_\pm \equiv \sqrt{ (\tilde{\omega}_m \pm \omega_m^\prime/2)^2 + \Delta_0^2} \  ,
\end{align}
which reduces to the normal state expression, Eq.~\eqref{Landaudampingdef}, for $\Delta_0 = 0$. We have suppressed the dependence on the quasiparticle weight $Z = 1/(1+\lambda) \simeq 1$.  To match
$\text{Im}\left[\Sigma_\text{bs}^R(\omega)\right]$ with the self-energy obtained using normal state propagators, we need $\Pi_{\text{ph}}$ for $v_F q \gg \Delta_0$. To extract the singular part of $\Pi_{\text{ph}}$ for real frequencies, we can take a suitable $\tilde{\omega}_m \rightarrow 0$ limit in the integrand, and also perform the analytical continuation $i\omega^\prime_m \rightarrow \omega^\prime_m$ before integration; this results in the correct real part for $\Pi_{\text{ph}}(\omega^\prime)$, and the imaginary part can then be restored from Kramers-Kronig relations. We have
\begin{align}
&\Pi_\text{ph}(\omega^\prime, v_F q) \simeq - \frac{\rho N}{v_F q} \frac{(\omega^\prime)^2}{2} \times  \\ & \notag \int_0^\Lambda  d\tilde \omega_m \frac{1}{\sqrt{(\Delta_0^2 - (\omega^\prime)^2/4)^2 + 2(\tilde\omega_m)^2 (\Delta_0^2 + (\omega^\prime)^2/4)} } \ ,
\end{align}
where $\Lambda$ is some arbitrary cutoff. This integral is singular for $\omega^\prime \rightarrow 2\Delta_0$, and we find \begin{align}
\Pi_{\text{ph}}(\omega^\prime, q) \simeq  \frac{\rho N}{v_F q} \Delta_0 \ln\left( \frac{ |\omega^\prime - 2\Delta_0|}{\Lambda} \right) .
\end{align}
Restoring the imaginary part to get a retarded function, we obtain
\begin{align}
\text{Im}[\Pi_{\text{ph}}^R(\omega^\prime, q)] \simeq - \pi \frac{\rho N }{v_F q} \Delta_0 \times \theta (\omega^\prime - 2\Delta_0) \ .
\end{align}
This is the expected behavior: since quasiparticles and holes are gapped with gap $\Delta_0$, the imaginary part of the polarization bubble should start at $2\Delta_0$. With this expression at hand, we can approximate
\begin{align}
\label{ImDRres}
\text{Im} \left[ D^R(\omega^\prime, \q) \right]   \simeq \frac{g\text{Im}\left[\Pi_{\text{ph}}^R(\omega^\prime, q)\right]}{\left(q^2 + M^2\right)^2},
\end{align}
which holds for $q >q_{\text{min}} \equiv  (g \rho N \Delta)/v_F M^2$, similar to the normal state case (compare Eq.~\eqref{qminApp}). Further, we have, for $\omega - \omega^\prime > 0$:
\begin{align} \notag
&\text{Im}\left[G^R(\omega - \omega^\prime, \k + \q) \right] = - \pi u_{\k+\q}^2 \delta\left( \omega - \omega^\prime - E_{\k+\q}
\right),   \\ & \ E_{\k + \q} = \sqrt{\Delta_0^2 + \xi_{\k + \q}^2}, \ u_{\k + \q}^2 = \frac{1}{2} \left( 1+ \frac{\xi_{\k +\q}}{E_{\k + \q}} \right) .  \label{ImGRres}
\end{align}
Inserting Eqs.\ \eqref{ImDRres}, \eqref{ImGRres} into Eq.\ \eqref{spectralrep}, we find, taking the $\omega^\prime$ integral:
\begin{align}
&\text{Im} \left[ \Sigma_{\text{bs}}^R(\omega) \right] = - \pi \frac{g^2 \rho N \Delta_0}{v_F} \int_{q> q_{\text{min}}}   \frac{d\q}{(2\pi)^2} \\ \notag & \theta\left( \omega - E_{\k+\q} - 2\Delta_0\right) \frac{u_{\k + \q}^2}{q(q^2 + M^2)^2} \ .
\end{align}
It is readily seen that the self-energy starts at $\omega = 3\Delta_0$. For $0< \omega - 3\Delta_0 \ll \Delta_0$, we can expand $E_{\k + \q} \simeq \Delta_0 + (v_F q \cos(\phi))^2/2\Delta_0$, and the integral is dominated by angles $\phi = \measuredangle(\k,\q) \simeq \pm \pi/2$. We can then write:
\begin{align} \notag
&\text{Im} \left[ \Sigma_{\text{bs}}^R(\omega) \right] = -  \frac{g^2 \rho N \Delta_0}{v_F} \frac{1}{4\pi} \int_{q_\text{min}} dq \frac{1}{(q^2 + M^2)} \notag \\& \notag \int dx \theta \left( \omega - 3\Delta_0 - (v_F q)^2/2\Delta_0 \times x^2  \right) = \\ & \notag  -  \frac{g^2 \rho N \Delta_0^{3/2}}{v_F^2} \frac{ \sqrt{(\omega - 3\Delta_0)}}{\sqrt{2}\pi} \theta(3 - \Delta_0)   \int_{q_\text{min}} dq \frac{1}{(q^2 + M^2)} \frac{1}{q}  \\  & =  - 2\sqrt{2} \frac{\Delta_0^2}{\omfl}  \lambda \ln\left(\frac{\omfl}{\Delta_0} \right)  \times \theta(\omega - 3\Delta_0) \sqrt{\frac{\omega - 3\Delta_0}{\Delta_0}} \ ,
\end{align}
as in {Eq.~\eqref{Imsigma3delta}}

\section{Self-energy in the superconducting state: $s$-wave contribution}
\label{swaveselfApp}

\begin{figure}
\centering
\includegraphics[width=.7\columnwidth]{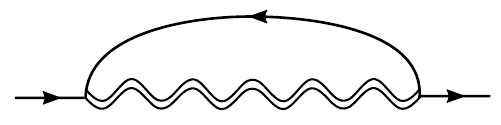}
\caption{Self-energy diagram which determines the $s$-wave part of the self-energy in the superconducting state. The double wavy line represents a Goldstone mode propagator, defined such that the vertex between it and the fermionic propagators is unity.}
\label{GSmodediag}
\end{figure}

The proper cutoff to the $s$-wave contribution in the normal state can be obtained by evaluating the scattering of a particle with a phase fluctuation of the order parameter (Goldstone mode), described by the diagram of Fig.\ \ref{GSmodediag}. It corresponds to
\begin{align}
\label{BssselfApp}
\Sigma_\text{bs,s}(\omega_m) = N\int_q G(\omega_m^\prime - \omega_m, \q - \k) B(\omega^\prime_m,\q) \ \ ,
\end{align}
Here, the $B$ is the Goldstone mode propagator, which can be obtained from the $s$-wave part of the backscattering amplitude, Eq.\ \eqref{remsum},
\begin{align}
\label{bsvertexlastApp}
\Gamma_\text{bs}  = -\frac{1}{\rho} \frac{\lamt}{1- \lamt \ln \left( \frac{\omfl}{v_F q \epsilon} \right)} \ ,
\end{align}
by replacing the logarithm with the $s$-wave piece of the particle-particle bubble $\Pi_{\text{pp}}(\omega_m^\prime,\q)/\rho$. In the superconducting state, $\Pi_\text{pp}(\omega_m^\prime,\q)$ corresponds to $GG + FF$:
 \begin{align}
\Pi_{\text{pp}}(\omega_m^\prime, \q ) &=  \int_p \left[ G(q-p) G(p) + F(q-p) F(p) \right] \ .
\end{align}
Let us first consider $\Pi_{\text{pp}}(0)$. It fulfills \begin{align}
\Pi_{\text{pp}}(0) \simeq \rho \ln\left(\omfl/\Delta_0 \right) = \rho/\lamt \ ,
\end{align}
where the last equation holds by definition of the $s$-wave gap. The energy-momentum dependent piece is only needed
for $\omega^\prime_m, v_F q \ll \Delta_0$. The integrands can then safely be expanded in $\omega^\prime_m, v_F q$, and the frequency integration can be extended to infinity. Neglecting all parts which are odd in the integration variables $\xi_\p, \tilde\omega_m, \cos(\phi)$ since they vanish upon integration, one obtains
\begin{widetext}
\begin{align}
\label{Cooperbubbleappres}
& \Pi_{\text{pp}}(\omega_m^\prime, q) - \Pi_{\text{pp}}(0) =   \rho \int \frac{d\phi}{2\pi} \frac{d\tilde{\omega}_m}{2\pi} d\xi_\p     \left\{ - \frac{\left[ (\omega_m^\prime)^2 + (v_F q \cos(\phi))^2 \right]}{\left((\tilde\omega_m)^2 + \xi_\p^2 + \Delta_0^2\right)^2} + \frac{\left[ 2(\omega_m^\prime)^2 (\tilde\omega_m)^2  + 2\xi_\p^2(v_F q \cos(\phi))^2 \right] }{\left((\tilde\omega_m)^2 + \xi_\p^2 + \Delta_0^2)\right)^3} \right\}  \\ & \notag   = -  \frac{\rho}{4\Delta_0^2} \left[ (\omega_m^\prime)^2 + \frac{1}{2}(v_F q )^2 \right] \ .
\end{align}
\end{widetext}
Diving this result by $\rho$ and inserting into Eq.\ \eqref{bsvertexlastApp} yields a Goldstone mode propagator
\begin{align}
B(\omega_m^\prime, \q) = - \frac{1}{\rho}\frac{4\Delta_0^2}{(\omega_m^\prime)^2 + \frac{1}{2}\left(v_F q \right)^2 } \ .
\end{align}
Inserting this into Eq.\ \eqref{BssselfApp}, we have
\begin{align}
\label{Sigma0scApp}
&\Sigma_\text{bs,s}(\omega_m) = \\ \notag &  -N \int_q  \left( \frac{u_{\q - \k}^2}{i(\omega_m^\prime - \omega_m) - E_{\q - \k}} + \frac{v_{\q - \k}^2}{i(\omega_m^\prime - \omega_m) + E_{\q - \k}}\right) \\ \notag &\quad \times \frac{1}{\rho}\frac{4\Delta_0^2}{(\omega_m^\prime)^2 + \frac{1}{2}\left(v_F q \right)^2 } \ ,  \quad v_{\q - \k}^2 = \frac{1}{2}\left(1-\frac{\xi_{\q-\k}}{E_{\q-\k}} \right) \ .
\end{align}

We perform the frequency integral, closing the contour such that we only encircle one bosonic pole for both the $u_{\q-\k}, v_{\q-\k}$ terms. This yields
\begin{align}
\label{swaveinterm}
&\Sigma_\text{bs,s}(\omega_m) =  - \frac{N  \Delta_0^2}{\sqrt{2}\pi^2 v_F \rho  }   \times \\ &   \int d\q \frac{1}{q} \left( \frac{u_{\q-\k}^2}{-i\omega_m - E_{\q-\k} - \frac{v_F q}{\sqrt{2}} } + \frac{v_{\q-\k}^2}{-i\omega_m + E_{\q-\k} + \frac{v_F q}{\sqrt{2}} }  \right) . \notag
\end{align}
We perform the analytical continuation and take the imaginary part:
\begin{align}
&\text{Im} \left[ \Sigma^R_\text{bs,s}(\omega) \right] =  \\  \notag &   - \frac{N \Delta_0^2}{\sqrt{2}\pi v_F \rho  }  \int d\q  \frac{1}{q}  \bigg[&& u_{\q-\k}^2 \delta\!\left(\omega + E_{\q - \k} + \frac{v_F q}{\sqrt{2}}\right)   \\ & \notag   &&\hspace{-1.2em}+   v_{\q-\k}^2  \delta\!\left(\omega - E_{\q - \k} - \frac{v_F q}{\sqrt{2}}\right)  \!\bigg] \ .
\end{align}
We focus on $\omega \gtrsim \Delta_0$, which comes from the part $\propto v_{\q - \k}$ and  $v_Fq \ll \Delta_0$. To leading order in $v_Fq/\Delta_0$, one can approximate $E_{\q-\k} \simeq \Delta_0$ and $v_{\q - \k} \simeq 1/2$, which gives
\begin{align} \notag
\text{Im} \left[ \Sigma_\text{bs,s}^R(\omega) \right] &\simeq - \frac{N \Delta_0^2}{\sqrt{2}v_F  \rho} \int_0^\infty dq \ \delta\!\left(\omega - \Delta_0 - \frac{v_F q}{\sqrt{2}}\right)  \\ &= - \pi  \frac{\epsilon^2}{\lambda}  \theta(\omega - \Delta_0) \frac{\Delta_0^2}{\omfl}   \notag ,
 \end{align}
 as stated in the main text.

\bibliographystyle{apsrev4-1}

\bibliography{ising_planar}

\end{document}